\documentclass[twocolumn]{aastex63}

\usepackage{natbib}
\usepackage{CJK}
\usepackage{longtable}
\usepackage{amsmath}

\makeatother

\submitjournal{ApJS}

\shorttitle{edge-on LSBG sample}
\shortauthors{He et al.}

\begin{document}
\begin{CJK*}{UTF8}{gkai}

\title{A Sample of Edge-on H{\sc{i}}-rich low-surface-brightness Galaxy Candidates in the 40\% ALFALFA Catalog}

\correspondingauthor{Hong Wu, Wei Du, Min He}
\email{Email: hwu@bao.ac.cn, wdu@bao.ac.cn, hemin13@live.com}

\author{Min He(何敏)}
\affiliation{Key Laboratory of Optical Astronomy,
       National Astronomical Observatories,
       Chinese Academy of Sciences,
       20A Datun Road, Chaoyang District,
       Beijing 100101, China}

\affiliation{School of Astronomy and Space Science,
             University of Chinese Academy of Sciences,
             19A Yuquan Road, Shijingshan District,
             Beijing, 100049, China}

\author{Hong Wu(吴宏)}
\affiliation{Key Laboratory of Optical Astronomy,
             National Astronomical Observatories,
             Chinese Academy of Sciences,
             20A Datun Road, Chaoyang District,
             Beijing 100101, China}

\affiliation{School of Astronomy and Space Science,
             University of Chinese Academy of Sciences,
             19A Yuquan Road, Shijingshan District,
             Beijing, 100049, China}

\author{Wei Du(杜薇)}
\affiliation{Key Laboratory of Optical Astronomy,
             National Astronomical Observatories,
             Chinese Academy of Sciences,
             20A Datun Road, Chaoyang District,
             Beijing 100101, China}

\author{He-yang liu(刘禾阳)}
\affiliation{Key Laboratory of Space Astronomy and Technology, 
             National Astronomical Observatories, 
             Chinese Academy of Sciences, 
             20A Datun Road, Chaoyang District, 
             Beijing 100101, China}

\affiliation{School of Astronomy and Space Science,
             University of Chinese Academy of Sciences,
             19A Yuquan Road, Shijingshan District,
             Beijing, 100049, China}

\author{Feng-jie Lei(雷凤杰)}
\affiliation{Key Laboratory of Optical Astronomy,
             National Astronomical Observatories, 
             Chinese Academy of Sciences, 
             20A Datun Road, Chaoyang District, 
             Beijing 100101, China}

\affiliation{School of Astronomy and Space Science,
             University of Chinese Academy of Sciences,
             19A Yuquan Road, Shijingshan District,
             Beijing, 100049, China}
                                
\author{Pin-song Zhao(赵品松)}
\affiliation{Key Laboratory of Optical Astronomy,
             National Astronomical Observatories, 
             Chinese Academy of Sciences, 
             20A Datun Road, Chaoyang District, 
             Beijing 100101, China}

\affiliation{School of Astronomy and Space Science,
             University of Chinese Academy of Sciences,
             19A Yuquan Road, Shijingshan District,
             Beijing, 100049, China}
             
\author{Bing-qing Zhang(张冰清)}
\affiliation{Key Laboratory of Optical Astronomy,
             National Astronomical Observatories, 
             Chinese Academy of Sciences, 
             20A Datun Road, Chaoyang District, 
             Beijing 100101, China}

\affiliation{School of Astronomy and Space Science,
             University of Chinese Academy of Sciences,
             19A Yuquan Road, Shijingshan District,
             Beijing, 100049, China}

\begin{abstract}

low-surface-brightness galaxies(LSBGs) are defined as galaxies that are fainter than dark night sky and are important for studying our universe. Particularly, edge-on galaxies are useful for the study of rotational velocity and dynamical properties of galaxies. Hence here we focus on searching for edge-on LSBGs. In order to find these edge-on dim galaxies, a series of effects caused by inclination, including the surface brightness profile, internal extinction, and scale length, have been corrected. In this work, we present a catalog of 281 edge-on LSBG candidates, which are selected from the cross-match between SDSS DR7 and the 40\% ALFALFA catalog. We also present the properties of these edge-on LSBG candidates including absolute magnitude, central surface brightness, $B-V$ color, scale length, and relative thickness. Our result suggests that the correction of inclination effects is very important for obtaining a complete sample of LSBGs.

\end{abstract}


\section{Introduction} 
\label{sec:introduction}

\defcitealias{Du2015}{Du+15}

low-surface-brightness galaxies (LSBGs), which was first introduced by \citet{Zwicky1957}, are galaxies that are fainter than the dark night sky \citep[e.g.,][]{Freeman1970, McGaugh1995}. LSBGs are commonly selected in the $B$ band with their central surface brightness below a certain threshold, which lies in the range of $21.5 - 23.0$ mag$\cdot$arcsec$^{-2}$ \citep[e.g.,][]{ONeil1997, Impey1997, Zhong2008, Du2015}. Apart from $B$ band, the central surface brightness in other bands, including $r$/$R$ band \citep[e.g.,][]{Courteau1996, Adami2006} and $K_{\rm s}$ band \citep[e.g.,][]{Jarrett1998, Jarrett2000a, Jarrett2000b, Monnier-Ragaigne2003a, Monnier-Ragaigne2003b, Monnier-Ragaigne2003c} are also used to search for LSBGs.\\

The LSBGs are of particular importance to the study of our universe since they may occupy a large fraction \citep[$30\% \sim 60\%$, e.g.,][]{McGaugh1995, McGaugh1996, Impey1997, ONeil2000a, Trachternach2006, Haberzettl2007, Martin2019} in the local galaxy population. Moreover, contrast to the high surface brightness galaxies (HSBGs), LSBGs present some different properties, including more extended shape \citep[e.g.,][]{de-Blok1996, de-Blok1997, de-Blok2001}, lower star formation rate \citep[e.g.,][]{van-der-Hulst1993, van-Zee1997, van-den-Hoek2000, Wyder2009, Schombert2011}, lower metallicity \citep[e.g.,][]{de-Blok1998a, de-Blok1998b, Kuzio-de-Naray2004}. Furthermore, the evolution of LSBGs and HSBGs seem to be different. Nevertheless, some other literature have also shown the similarities in terms of dark matter fraction, star formation, chemical enrichment histories, and progenitor population between LSBGs and HSBGs\citep[e.g.,][]{Zwaan1995, Gao2010, Liang2010, Martin2019}. Studies on LSBGs and HSBGs can help to gain our understanding of galaxy formation and evolution. And increasing sample size of LSBGs can provide unique opportunities to study galaxies at the low-surface-brightness end.\\

With the development of facilities, amounts of LSBG samples have been identified in last decades \citep[e.g.,][]{Schombert1992, McGaugh1995, de-Blok1995, Impey1996, Zhong2008, Du2015, Williams2016, Du2019}. In 2011, the Arecibo Legacy Fast ALFA (ALFALFA) Survey provided a 40\% catalog ($\alpha.40$ catalog) of extragalactic H{\sc{i}} line sources\footnote{\url{http://egg.astro.cornell.edu/alfalfa/data/index.php}}. This catalog is practical to search for LSBGs, since LSBGs are considered to harbor rich HI gas. Using the $\alpha.40$ catalog, \citet[][hereafter Du+15]{Du2015} has found 1129 LSBGs, which are all non-edge-on galaxies.\\

There are many advantages to studying edge-on galaxies. The vertical orientation lead to higher surface brightness, and easier detection of fainter disk galaxies. In addition, their spindly morphology is suitable for spectroscopic observations and studies on the stellar populations of galaxies, the galaxy structure, and the contribution of dark matter to galaxies \citep[e.g.,][and references therein]{van-der-Kruit1981, Zasov1991, Grijs1998, Zheng1999, Wu2002, Du2017, Bizyaev2017}. In this paper, we aim at constructing a sample of edge-on LSBGs from the $\alpha.40$ catalog, which can be a good complement to the non-edge-on LSBGs sample \citepalias{Du2015}.\\

In section \ref{sec:data-sample}, we briefly introduce data analyses and parent sample. Section \ref{sec:CSB} describes the method of correcting the central surface brightness from edge-on phase to face-on phase. The sample and the properties of the edge-on LSBG candidates sample are presented in Section \ref{sec:sample-prop}, and discussions on uncertainties, LSBG fraction and comparison with HSBGs are shown in Section \ref{sec:discussion}, followed by a summary in the Section \ref{sec:summary}.\\

\section{Data and Parent Sample} \label{sec:data-sample}
\subsection{Data} \label{subsec:data}

Our LSBGs sample is selected from the optical-H{\sc{i}} cross-match from \citet{Haynes2011}. This cross-matching catalog is produced by the ALFALFA survey, which is a H{\sc{i}} survey with a sky of coverage of 7000 deg$^2$. The $\alpha.40$ catalog we used covers 2800 deg$^2$, including the ``spring'' region, $07^h30^m < R.A. < 16^h30^m$, $+04\arcdeg < decl. < +16\arcdeg $, and $+24\arcdeg < Dec. < +14\arcdeg $; and the ``fall'' region, $22^h < R.A. < 03^h$, $+14\arcdeg < Dec. < +16\arcdeg $, and $+24\arcdeg < Dec. < +32\arcdeg $.\\

There are 15,855 sources in the $\alpha.40$ catalog, 15,041 of which are extragalactic objects. Since the Sloan Digital Sky Survey (SDSS) data has a widely coverage and overlap with ALFALFA, 12,468 of these extragalactic sources have been found optical counterparts from the Sloan Digital Sky Survey Data Release 7 \citep[SDSS DR7][]{Haynes2011}\footnote{\url{http://classic.sdss.org/dr7/algorithms/dataProcessing.html}}.\\ 

\subsection{Data reduction} \label{subsec:reduction}

The SDSS DR7 pipeline tends to overestimate the sky background, which will result in an underestimation of galaxy luminosity. This effect is particularly significant for faint galaxies \citep{Lauer2007, Liu2008, Hyde2009, He2013, Du2015}. In order to alleviate this deviation, we have used a more adaptive measurement to estimate the sky background \citep{Zheng1999, Wu2002, Du2015} of SDSS fpC-images for the 12,468 cross-matched galaxies in the work of \citetalias{Du2015}.\\

An eight-pixel FWHM Gaussian Function is used to smooth the fpC images, which can enhance the optical shape of sources. After getting the smooth images, we detect sources by using SExtractor. With adjusting the parameters in ``Extraction section'' of SExtractor configuration file, we can detect the existing objects for each image. After getting a segmentation image about all the objects detected in the previous smooth image, we subtract these detected objects from fpC-image. Then, a low-order least square polynomial has been used to fit each row and each column of this object-subtracted image, and the masked pixels of this image are replaced with fitted values. We average the row-fitted values and column-fitted values of each pixels, and smooth the averaged image by 31$\times$31 pixels to obtain the finally sky image \citepalias{Du2015}. \\

Next, we use SExtractor again to photometry the target object after subtracting the fitting sky background from fpC-image. Among the AUTO, ISO and PETRO photometry modes, we choose the fitting result of AUTO mode which is photometry from the Kron flexible elliptical aperture \citep{Bertin1997}. From the output file, we can obtain the flux of the target galaxy in ADU unit, which can be transformed to magnitude by,
\begin{equation}
\rm mag = -2.5\log_{10}(\frac{counts}{exptime})-(aa+kk\cdot airmass),
\end{equation}
where `aa' is the zero point of fpC-image, and `kk' is the atmosphere extinction coefficient. `aa', `kk' and `airmass' can be gotten from SDSS drField*.fits files. The exposure time (exptime) here is 53.91 seconds for the SDSS DR7 photometry images. As many researches of LSBGs are base on $B$ band, we calculated the $B$-band magnitude by $B=g+0.47(g-r)+0.17$ \citep{Smith2002}.\\ 

\subsection{Parent Sample} \label{subsec:p-sample}

After photometry, the GALFIT software is chosen to fit the surface brightness profile of galaxies. By using the exponential model, we can obtain the axis ratio of these galaxies. Both $g$-band and $r$-band images are used for this fitting. Since \citetalias{Du2015} has selected a non-edge-on LSBGs sample by $b/a > 0.3$, we select the preliminary edge-on galaxies, whose axis ratios are $b/a \leq 0.3$ in $g$ band or $r$ band, to complement the non-edge-on sample. Besides that, \citet{Reshetnikov2019} have also selected a sample of edge-on galaxies from the Hubble Ultra Deep Field, and the mean ratio of the vertical and radial exponential disk scales of these edge-on galaxies is $0.25 \pm 0.07$, which suggests that our selection of $b/a \leq 0.3$ is reasonable. Both the $b/a$ values from GALFIT and the magnitudes from SExtractor are obtained directly from the work of \citetalias{Du2015}.\\

There are 2500 galaxies in accordance with this criterion of $b/a \leq 0.3$. We then visually check the SDSS $g$-,$r$-,$i$-band combined images of these galaxies independently by three persons, to see whether the galaxy appears to be a real edge-on galaxy. 1670 galaxies are considered to be edge-on galaxies, in which eight galaxies are removed because their images are not completely within the image frame.\\

\begin{figure*}[htb!]
\gridline{\fig{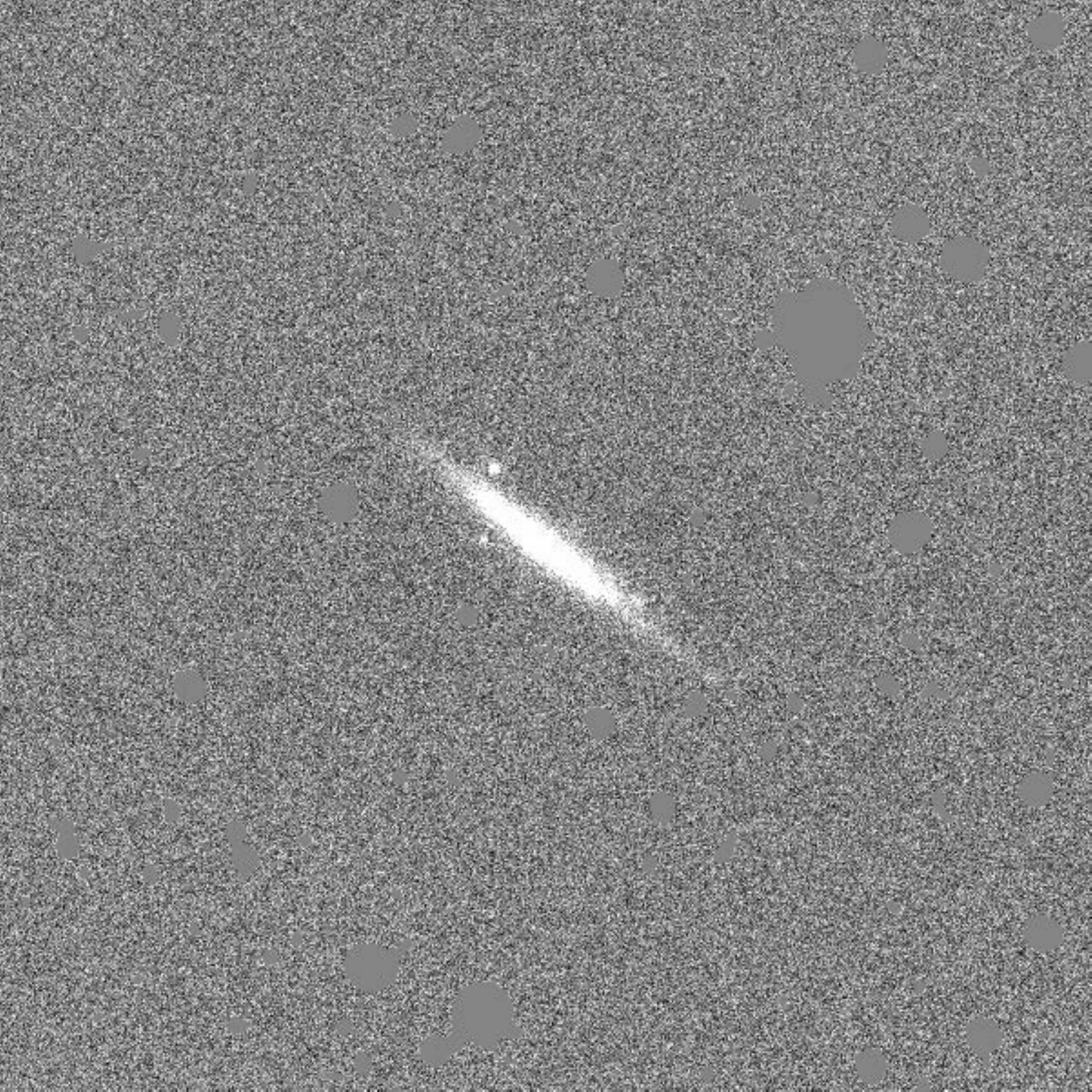}{0.3\textwidth}{fpC-image}
          \fig{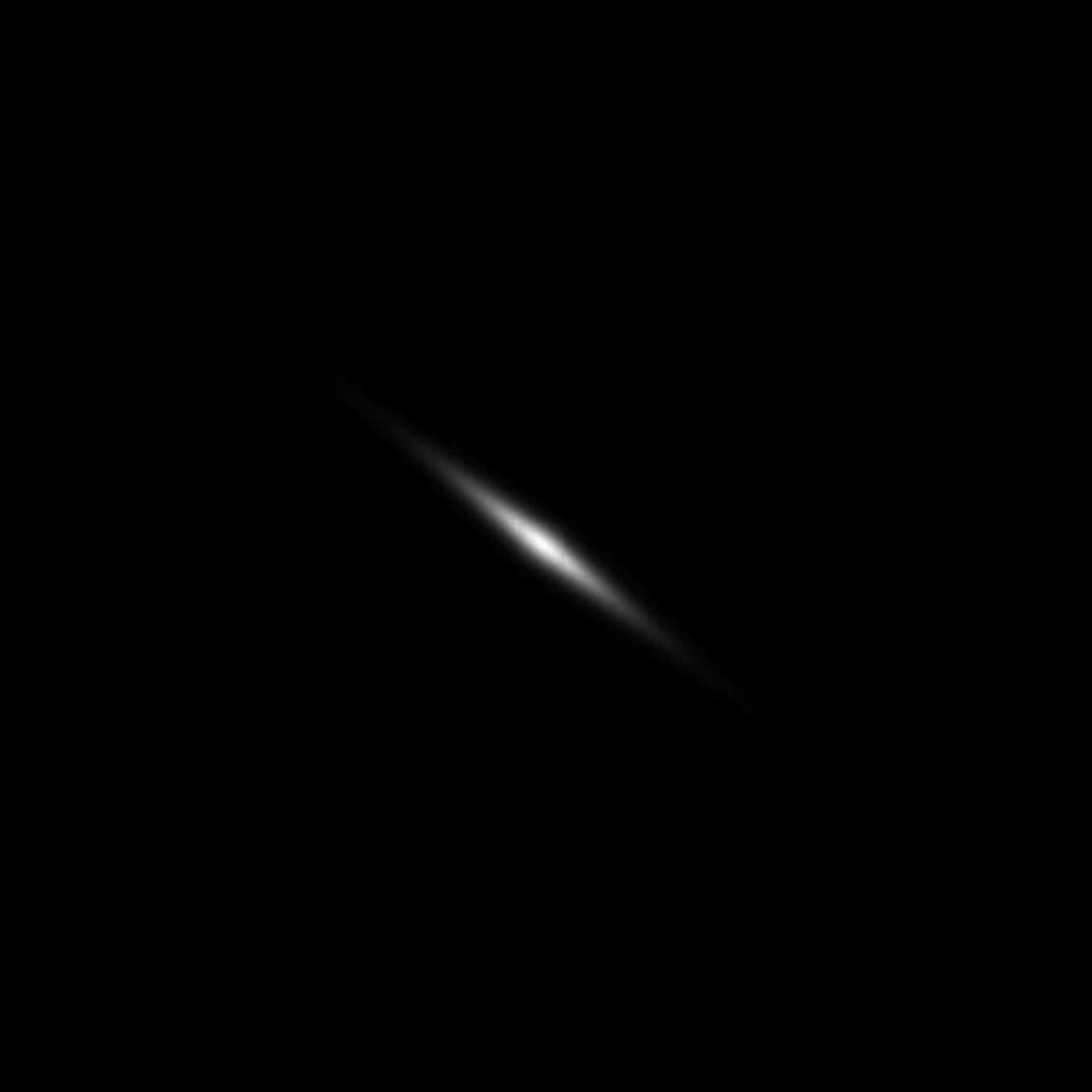}{0.3\textwidth}{edge-on-disk model}
          \fig{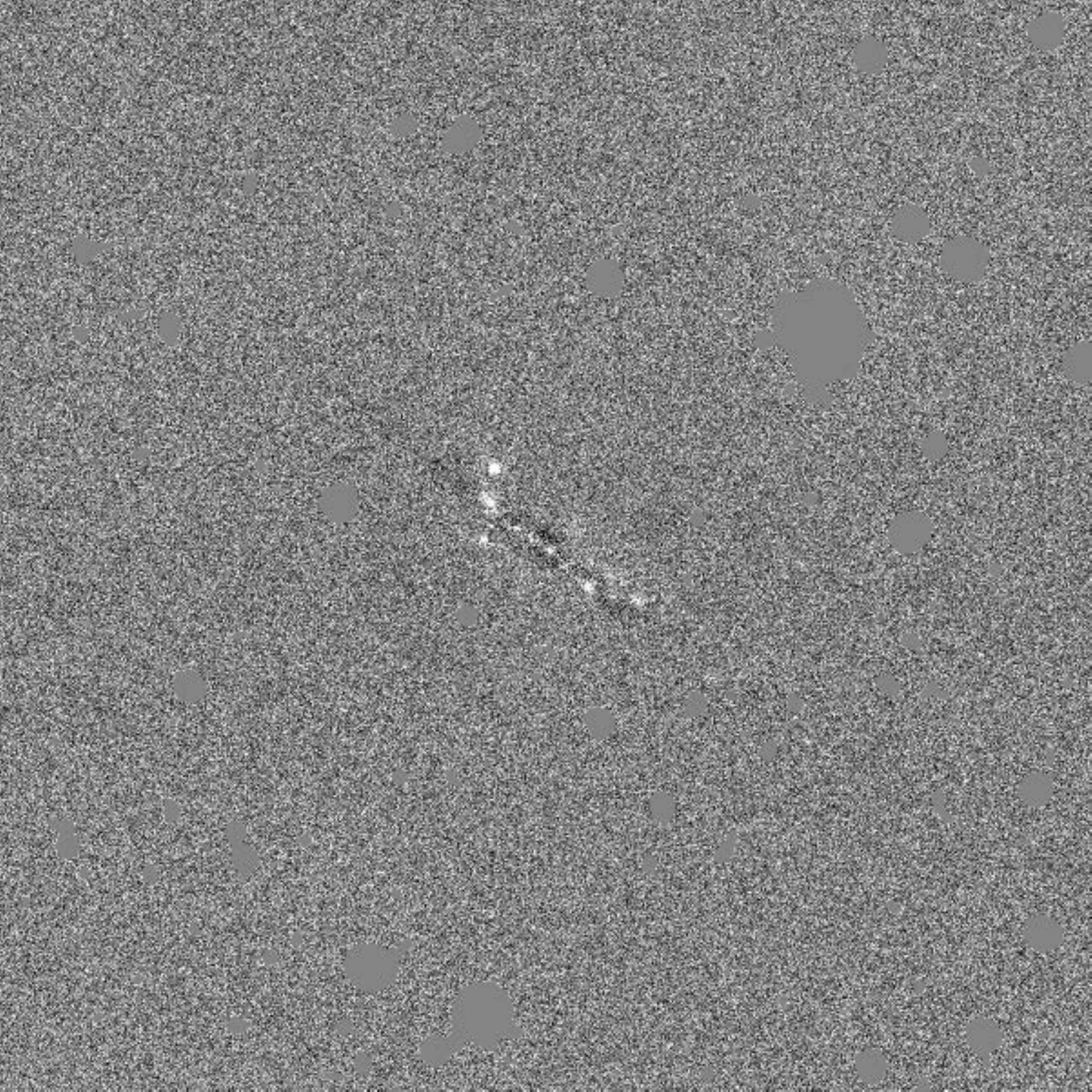}{0.3\textwidth}{residual image}
          }
\caption{Edge-on disk model fitting to the galaxy AGC 2221. These images are the fpC-image with sky subtraction and bright star removal, the model image and the residual image, from left to right respectively.
\label{fig:galfit}}
\end{figure*}

For obtaining the information of scale heights and better fitting for these galaxies, the edge-on exponential model,
\begin{equation}
\tilde{\mu}(R,h) = \tilde{\mu}_{\rm 0,edge}(\frac{R}{r_{\rm s}})K_{1}(\frac{R}{r_{\rm s}}){\rm sech}^2(\frac{h}{h_{\rm s}}),\label{eq:edge-on-function}
\end{equation}
is chosen to fit these edge-on galaxies. Since we have subtracted the sky background in Section 2, the sky background of the input image is fixed. And the fit-results will not produce significant differences, which are less than 1\%, from the fit-results produced by adding a free sky model for original fpC-image. Except the sky background, the bright objects around these edge-on galaxies have also been removed. The input PSFs used for fitting can be obtained from the psField-* files, which can be downloaded from the SDSS website\footnote{\url {http://classic.sdss.org/dr7/algorithms/dataProcessing.html}}.\\

One note is that the profile fitting for edge-on galaxies is extremely sensitive to the initial parameters, including the central surface brightness $\mu_{\rm 0,edge}$, scale length $r_{\rm s}$, scale height $h_{\rm s}$ and positional angle ${\rm PA}$. It is difficult to obtain the best-fit result if the initial values are far away from the best-fit ones. Thus, we give a range for each parameter according to the SExtractor outputs, such as $10\sim 25\rm\ mag\cdot arcsec^{-2}$ for $\mu_{0,edge}$, $2\rm\  pixels \sim 1.5\times \rm r_{s,SExtractor}$ for $r_{s}$, $1\rm\ pixel \sim 0.5\times r_{s}$ for $h_{s}$, and the PA is the output of SExtractor. Then we have made a 100 times loop to fit, in each loop we have selected the initial values randomly from these ranges. We stop the loop until the GALFIT outputs a good result successfully (a good result means that there are not star signs in the result, such as “*0.01*”). As the description in the work of \citet{He2019}, we have tested the scatter of all the good fitting results from a 100 times loop for one edge-on galaxy, and the output results are convergent. So, once this fitting outputs a good result in a loop , the result can be treated as the best-fit result we need. Finally, 1575 galaxies have been fitted successfully, and used as the parent sample. A fitting example of AGC 2221 is displayed in Figure \ref{fig:galfit}.\\

\section{Method of central surface brightness correction}
\label{sec:CSB}

To match the criterion of face-on LSBGs, two correction are most critical for central surface brightness. One is that the correction of surface brightness profile from edge-on to face-on. This could be fixed using theoretical models. The other is that the correction of internal extinction and scale length, which would alter along with the inclinations. This could be done by empirical relationships.\\

\subsection{Model correction of central surface brightness}
\label{subsec:model-corr}

Generally, low-surface-brightness galaxies are objects whose face-on disk central surface brightness are fainter than the night sky by at least one magnitude \citep{Freeman1970}. However, the measurement of the central surface brightness for a disk galaxy is subject to the line of sight. As mentioned in \cite{van-der-Kruit1981, Giovanelli1995, He2019}, for a face-on disk galaxy, its surface brightness profile can be described using
\begin{equation}
\tilde{\mu}(R) = \tilde{\mu}_{\rm 0,face} e^{-\frac{R}{r_{\rm s}}},
\end{equation}
where, $\tilde{\mu}_{\rm 0,face}$ is the observed central surface brightness when the galaxy is face-on,
\begin{equation}
\tilde{\mu}_{\rm 0,face}=2h_{\rm s}\rho_{0}.
\end{equation}
While for an edge-on galaxy, its surface brightness profile becomes Equation \ref{eq:edge-on-function},
\begin{equation}
\tilde{\mu}(R,h) = \tilde{\mu}_{\rm 0,edge}(\frac{R}{r_{\rm s}})K_{1}(\frac{R}{r_{\rm s}}){\rm sech}^2(\frac{h}{h_{\rm s}}),\tag{\ref{eq:edge-on-function}}
\end{equation}
and
\begin{equation}
\tilde{\mu}_{\rm 0,edge} = 2r_{\rm s}\rho_{0}.
\end{equation}
In these functions, $\rho_{0}$ is central luminosity density, $\tilde{\mu}(R)$ and $\tilde{\mu}(R,h)$ are surface brightness, and $\tilde{\mu}_{\rm 0,face}$ and $\tilde{\mu}_{\rm 0,edge}$ are the central surface brightness; $R$ and $h$ is the radial and vertical distance from the center, $r_{\rm s}$ and $h_{\rm s}$ are scale length and scale height of galaxy, respectively, and $K_{1}$ is the modified Bessel function. Therefore the relationship between central surface brightness of face-on and edge-on galaxies can be described using
\begin{equation}
\tilde{\mu}_{\rm 0,face}=\tilde{\mu}_{\rm 0,edge} \times \frac{h_{\rm s}}{r_{\rm s}},
\end{equation}
and when expressed in magnitude system, the above equation becomes:
\begin{equation}
\mu_{\rm 0,face}=\mu_{\rm 0,edge}-2.5\log_{10}(h_{\rm s}/r_{\rm s})\label{eq:correct-mu-simple}
\end{equation}
Obviously, the central surface brightness of a disk galaxy will become brighter if it's orientation change from face-on to edge-on, because its scale length is larger than scale height. And the difference of central surface brightness will be more significant in the case of a super-thin galaxy. Hence, the observed central surface brightness of an edge-on galaxy should be converted into face-on phase to keep in consistence with face-on galaxy.\\

\subsection{Correction of Internal Extinction and scale length}
\label{subsec:extinction}

Internal extinction can weaken the observed magnitudes, and it is particularly significant for edge-on galaxies. Therefore, in general, just by simply $-2.5\ log_{10} (h_{\rm s}/r_{\rm s})$ can not correct the central surface brightness of edge-on galaxies very well. Unfortunately, it is difficult to obtain a theoretical relationship between the internal extinction of face-on and edge-on galaxies. Here we try to explore the possible correlation between the internal extinction of face-on and edge-on galaxies empirically. \\

\begin{figure}[htb!]
\plotone{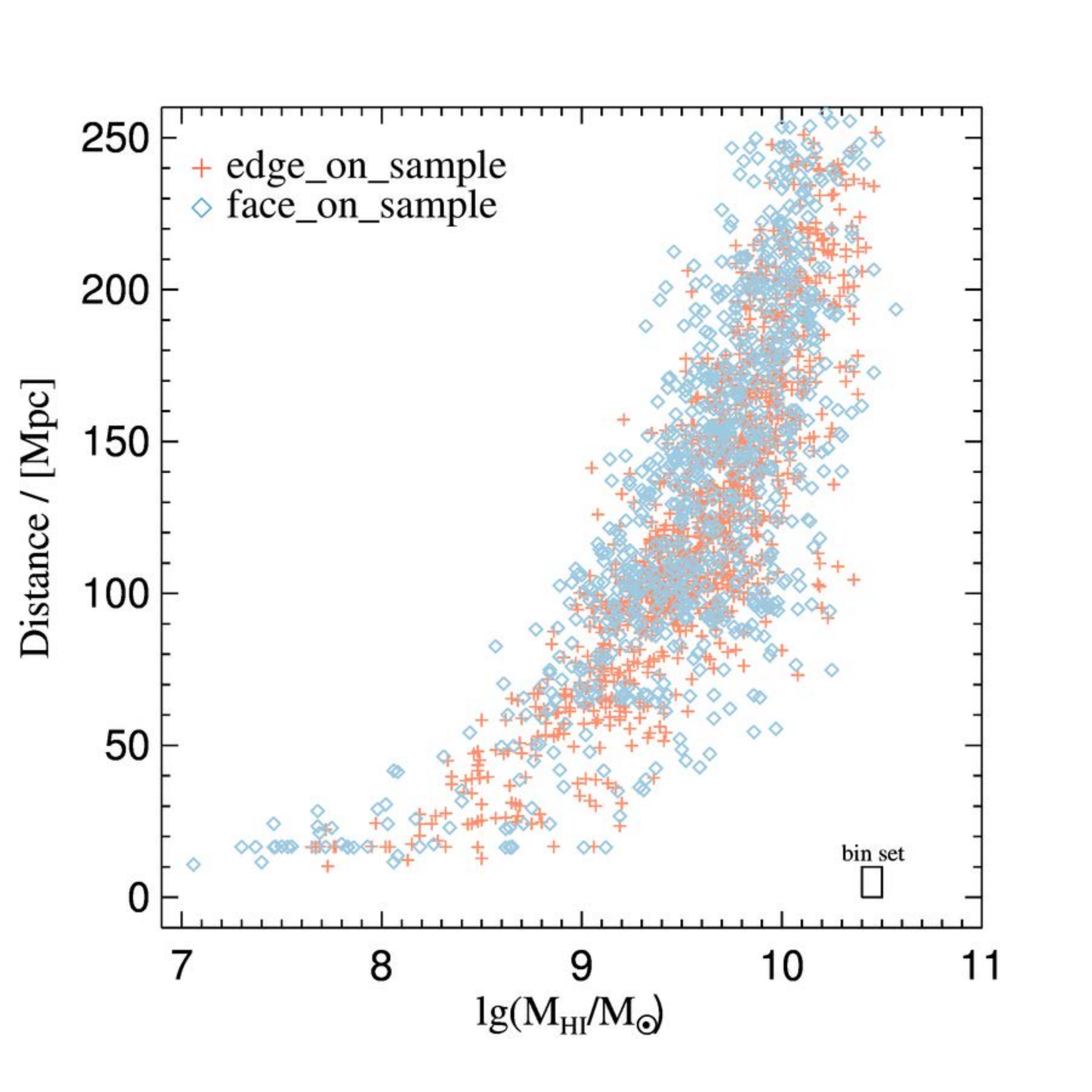}
\caption{The distributions of edge-on and face-on galaxies in $M_{\rm HI}$--Distance plot. Red plus and blue diamond symbols are sources from edge-on and face-on sample respectively. The edge-on and face-on samples are separated into several bins, and the bin size was presented in the lower right corner of this figure. The H{\sc{i}} mass is separated with a bin of 0.1 $\log_{10}(M_{\odot})$ and Distance is separated with a bin of 10 Mpc.\label{fig:bin_set}}
\end{figure}

To achieve this goal, two samples are selected from the $\alpha.40$ catalog, one is our edge-on galaxies, and the other is face-on galaxies. The selection criteria are as follows:\\
1. Face-on galaxy sample are selected by $b/a \geq 0.8$ \citep{Yoshino2015}.\\
2. The galaxies with unreasonable fitting results, whose $\chi^2$ are $\geq 3\sigma$, are rejected.\\
3. The galaxies whose scale lengths in both $g$ and $r$ band are too small ($r_{\rm s} < 2 \arcsec$, near the seeing of SDSS image) are also removed.\\
4. Irregular and interacting galaxies are rejected.\\
At last, 1013 edge-on galaxies and 907 face-on galaxies are obtained.\\

The following step is to match these two samples with common properties. Since all these galaxies are also detected by ALFALFA, in addition to the optical information, we can also obtain the H{\sc{i}} information of these galaxies, and some information of H{\sc{i}} may not be significantly affected by inclination. We roughly assume that galaxies tend to have similar statistical properties if they have similar H{\sc{i}} mass and are at roughly same distance and the H{\sc{i}} mass and distance are independent of the inclination. Hence the statistical difference in the observed magnitudes between edge-on and face-on galaxies with similar HI masses and distances are mainly due to their internal extinction.\\

According to this assumption, both edge-on and face-on galaxy samples are divided into several bins, the sizes of each bin are 0.1 $\log_{10}(M_{\odot})$ and 10 Mpc distance, as shown in Figure \ref{fig:bin_set}. Then, we calculate the mean value of each parameter of all the galaxies in the same bin for edge-on sample and face-on sample. Finally, we match the mean values in the same bin to construct a face-on-edge-on matched catalog.\\

\begin{figure*}[htb!]
\begin{center}
\gridline{\fig{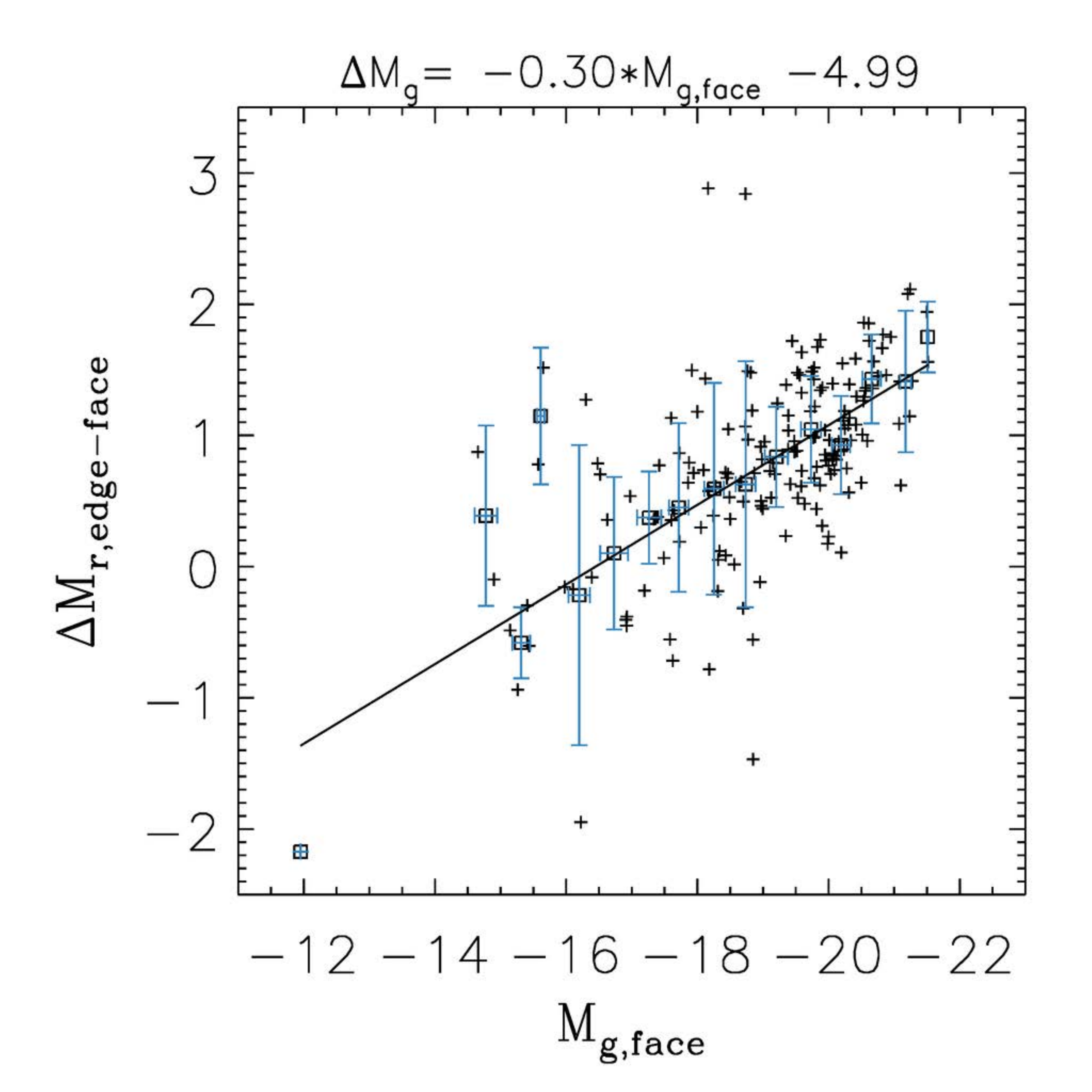}{0.33\textwidth}{(a)}
          \fig{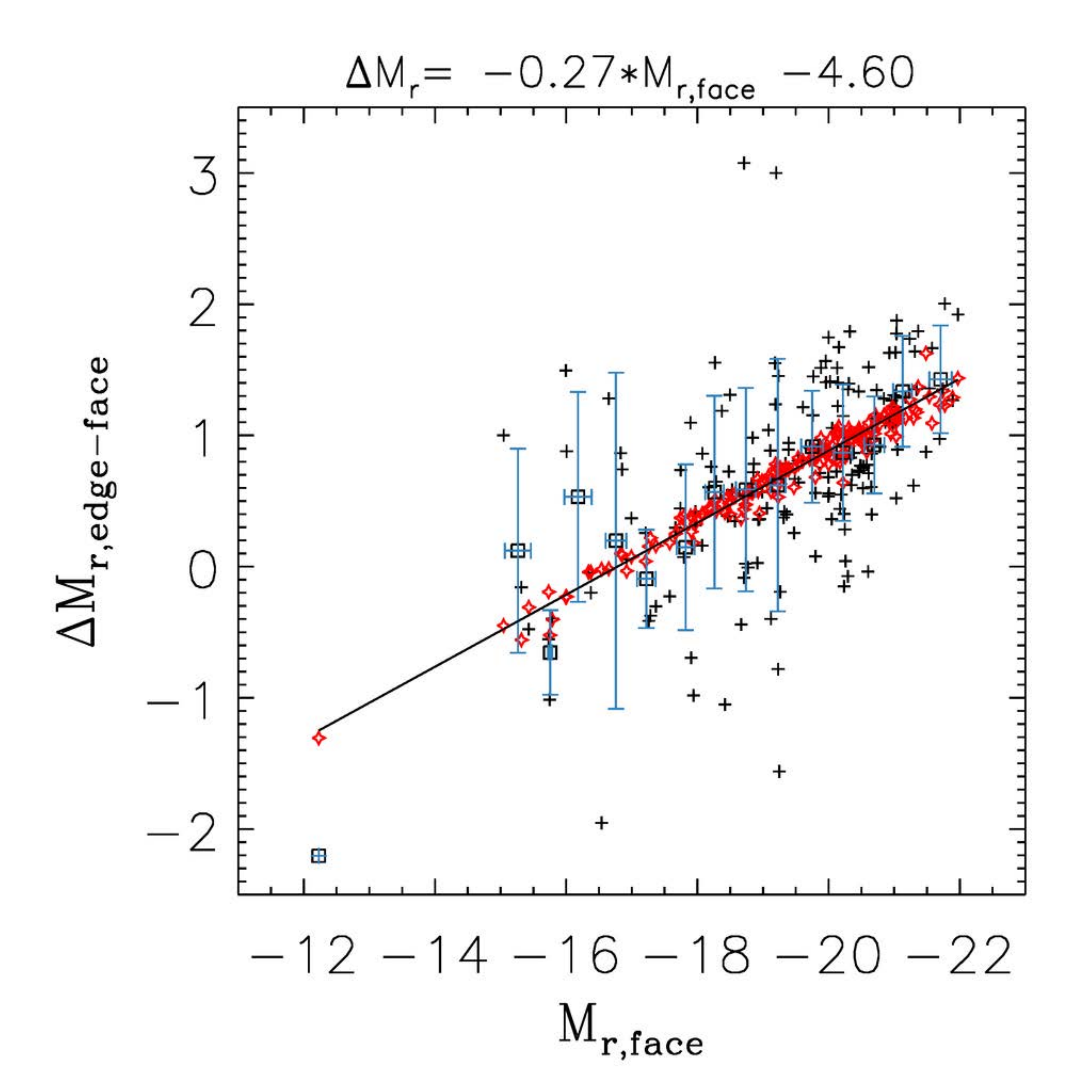}{0.33\textwidth}{(b)}
          \fig{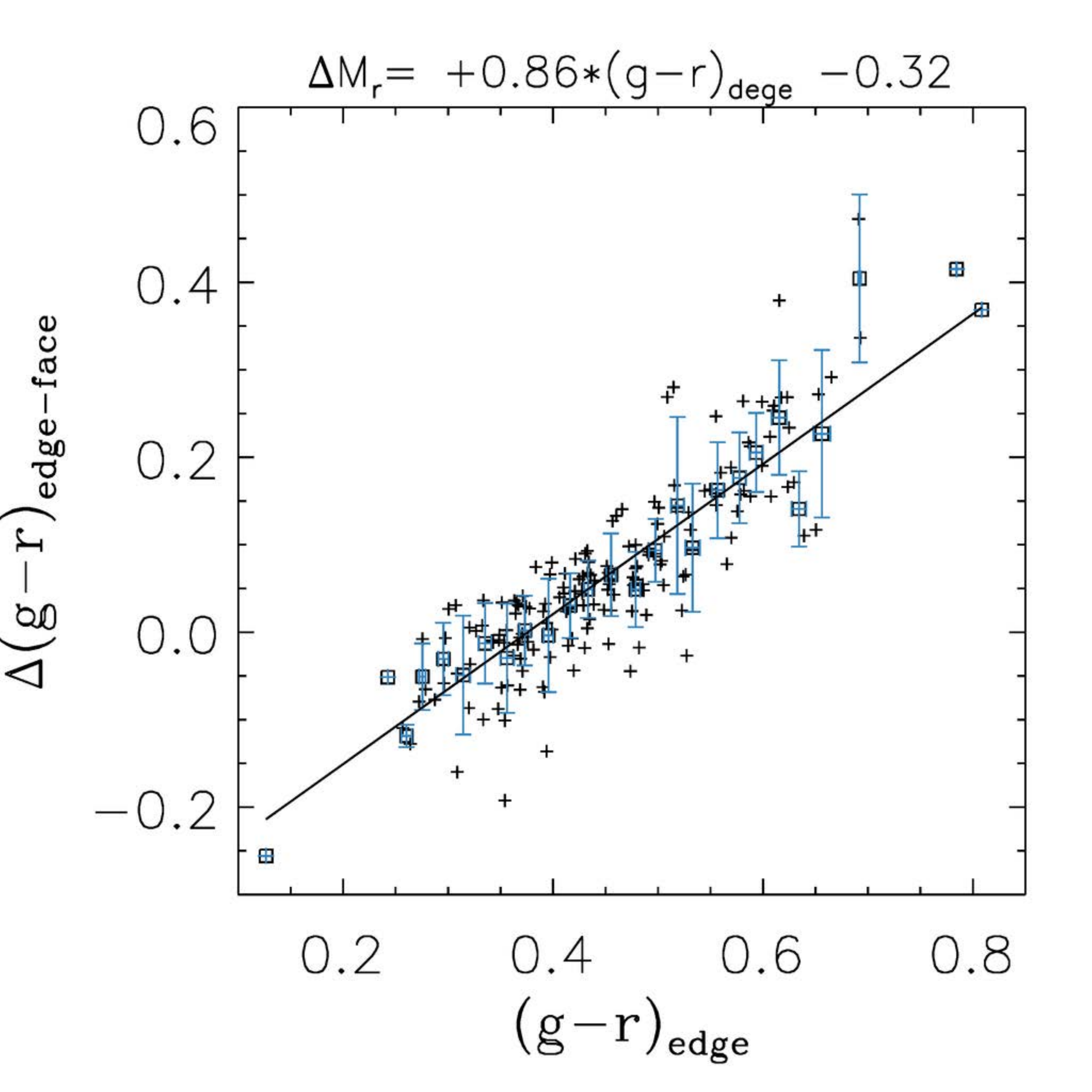}{0.33\textwidth}{(c)}
          }
\gridline{\fig{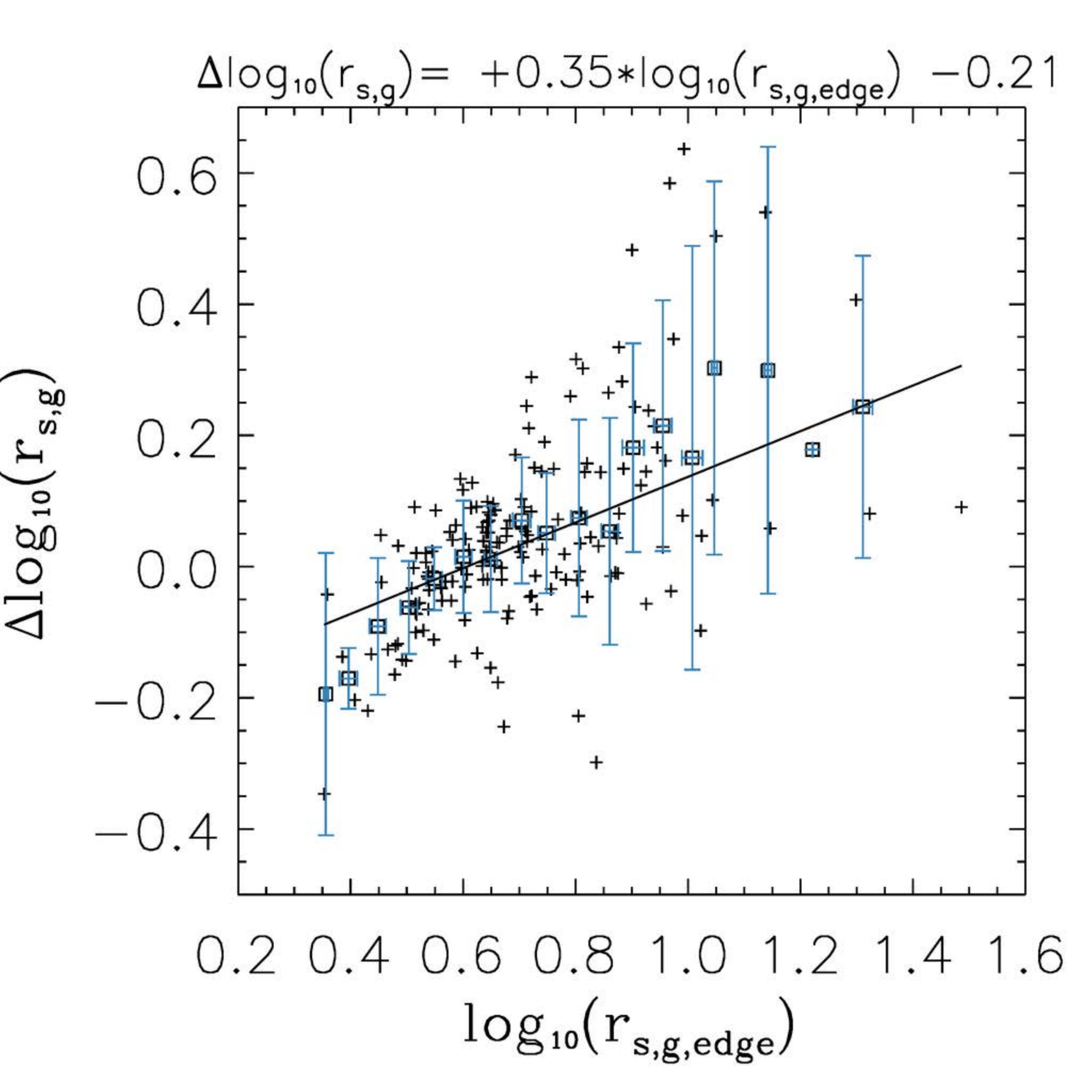}{0.33\textwidth}{(d)}
          \fig{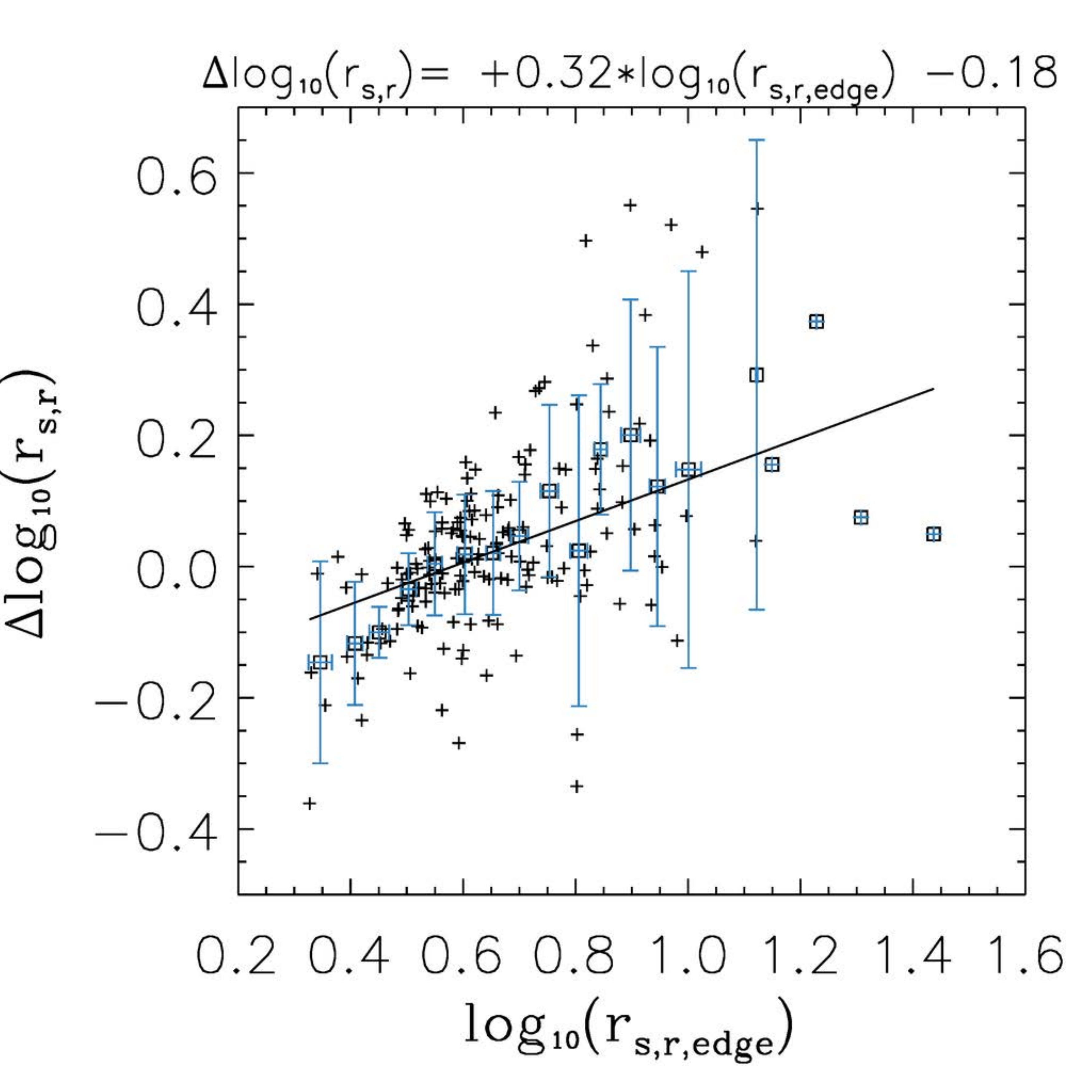}{0.33\textwidth}{(e)}
          }
\caption{Top panels: The relationships of $M_{\rm face}$ vs. $\Delta M$ for both $g$-band and $r$-band (panels a-b), and $(g-r)_{\rm edge}$ vs. $\Delta (g-r)_{\rm edge-face}$(panel c); Bottom panels: The relationships of $\Delta log_{10}(r_{s})$ vs. $log_{10}(r_{s,edge})$ for $g$/$r$-band (panels d-e). $M_{\rm face}$ is the face-on absolute magnitude, $\Delta M$ is the difference of absolute magnitude obtained by subtracting $M_{\rm face}$ from $M_{\rm edge}$; $(g-r)_{\rm edge}$ is the edge-on $g-r$ color, $\Delta (g-r)_{\rm edge-face}$ is the difference of $g-r$; $log_{10}(r_{s,edge})$ is the logarithm of the scale length of edge-on galaxies in a unit of arcsec to base 10, and $\Delta log_{10}(r_{s})$ is the difference of $log_{10}(r_{s})$ between edge-on and face-on galaxies. The black plus symbols in the figures are the data of the matched catalog used for fitting, and the fitting relationships obtained by robust linear fit are represented by lines. Squares are the mean values of each bin of x-axis, and are distributed near the fitting line. Red asterisks denote the results of $r$ band absolute magnitude calculated with the fitting results of $g$-band and $g-r$ color, and they are consistent with the fitting results of $r$-band, which indicates that our fitting results of the relationships between $\Delta M$ and $M_{\rm face}$ may be reliable.
\label{fig:fit-result}}
\end{center}
\end{figure*}

\begin{figure*}[htb!]
\begin{center}
\gridline{\fig{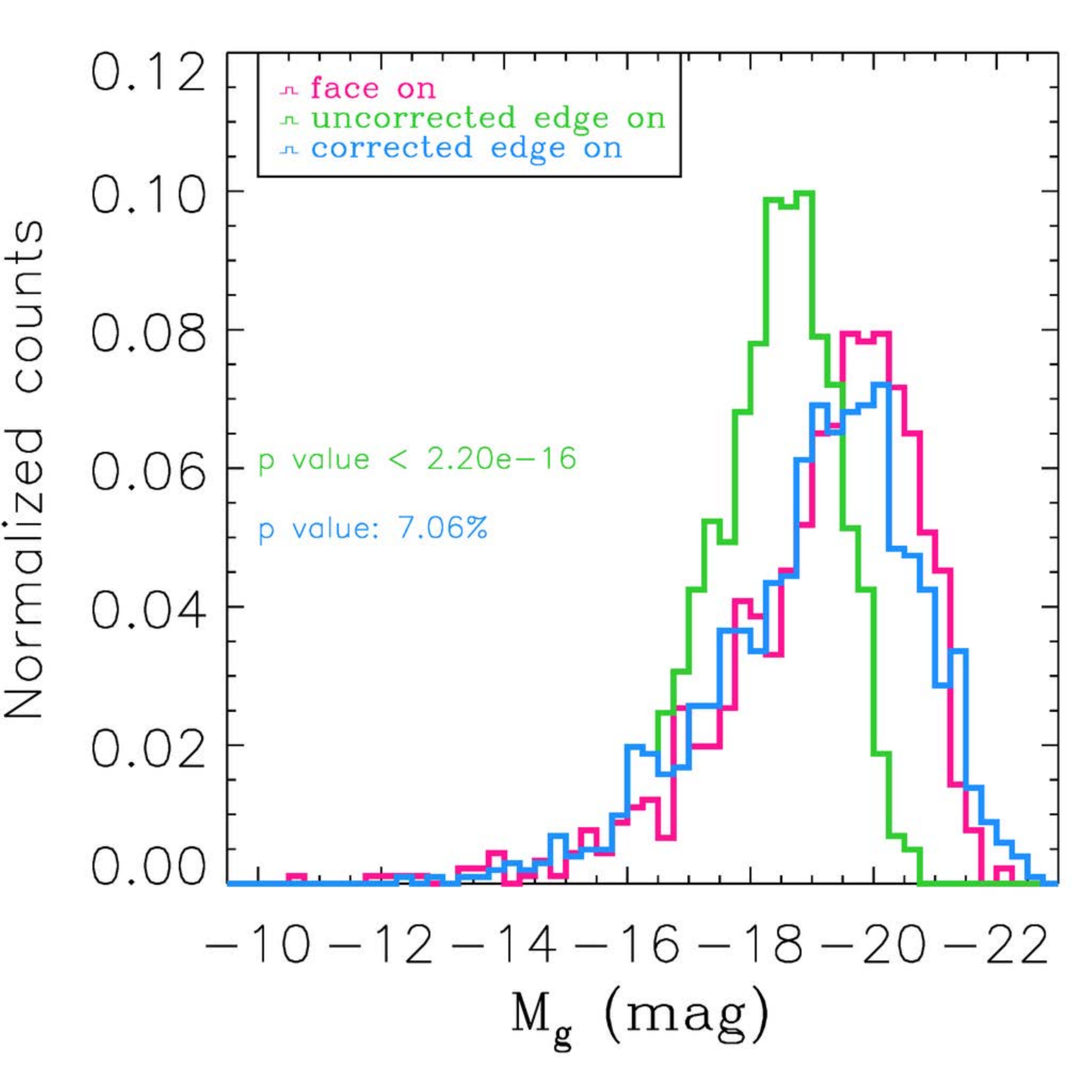}{0.33\textwidth}{(a)}
          \fig{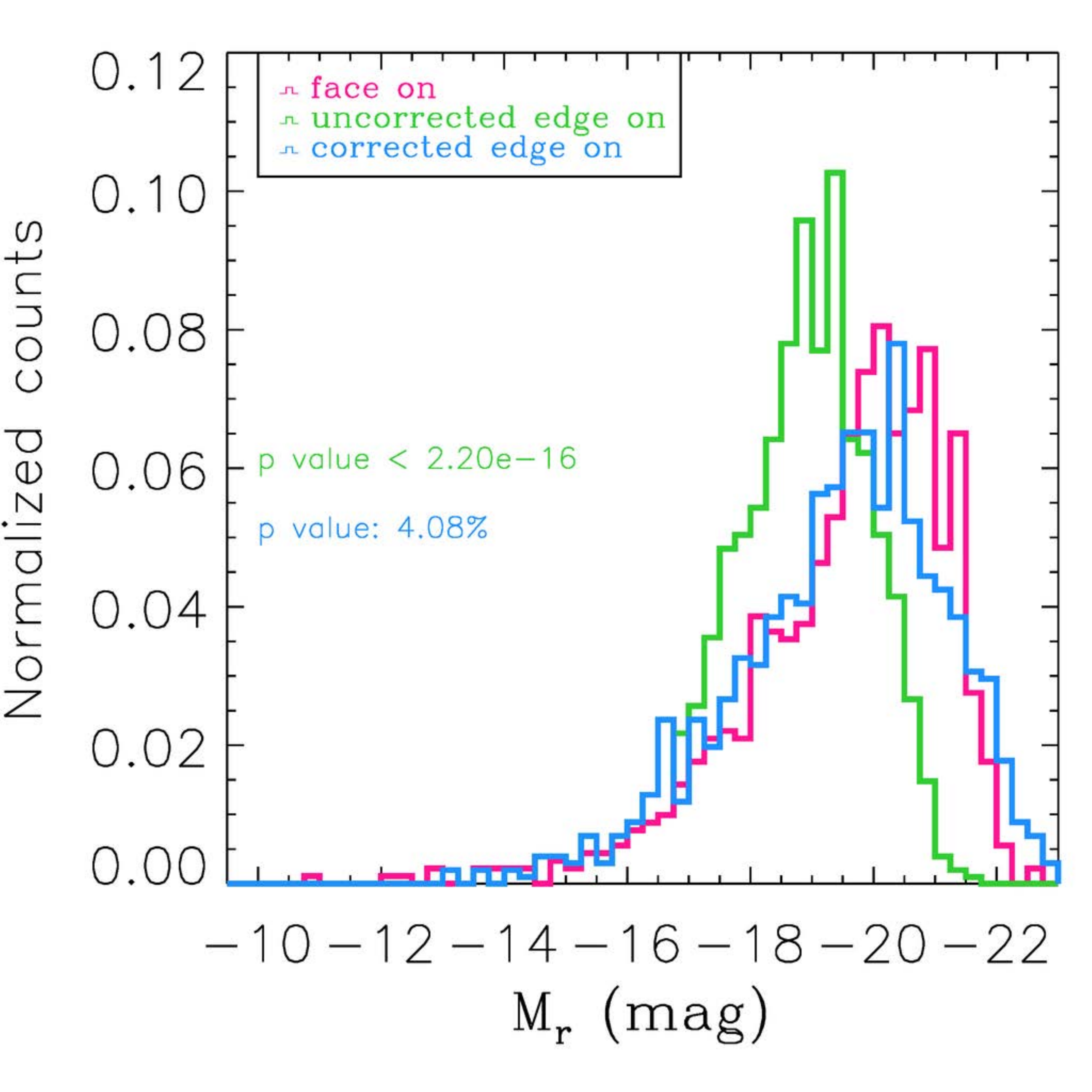}{0.33\textwidth}{(b)}
          \fig{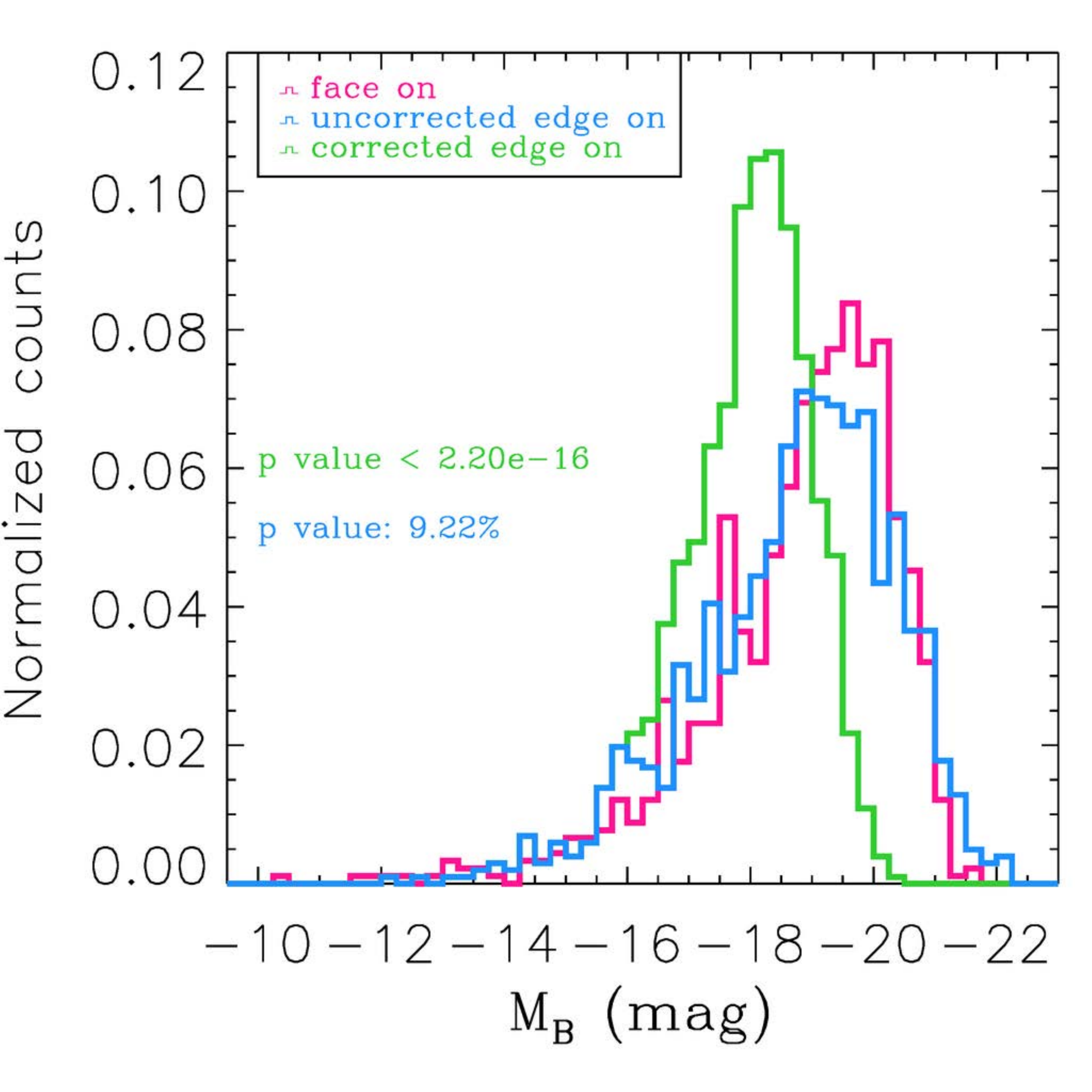}{0.33\textwidth}{(c)}
          }
\vspace{-0.5cm}
\gridline{\fig{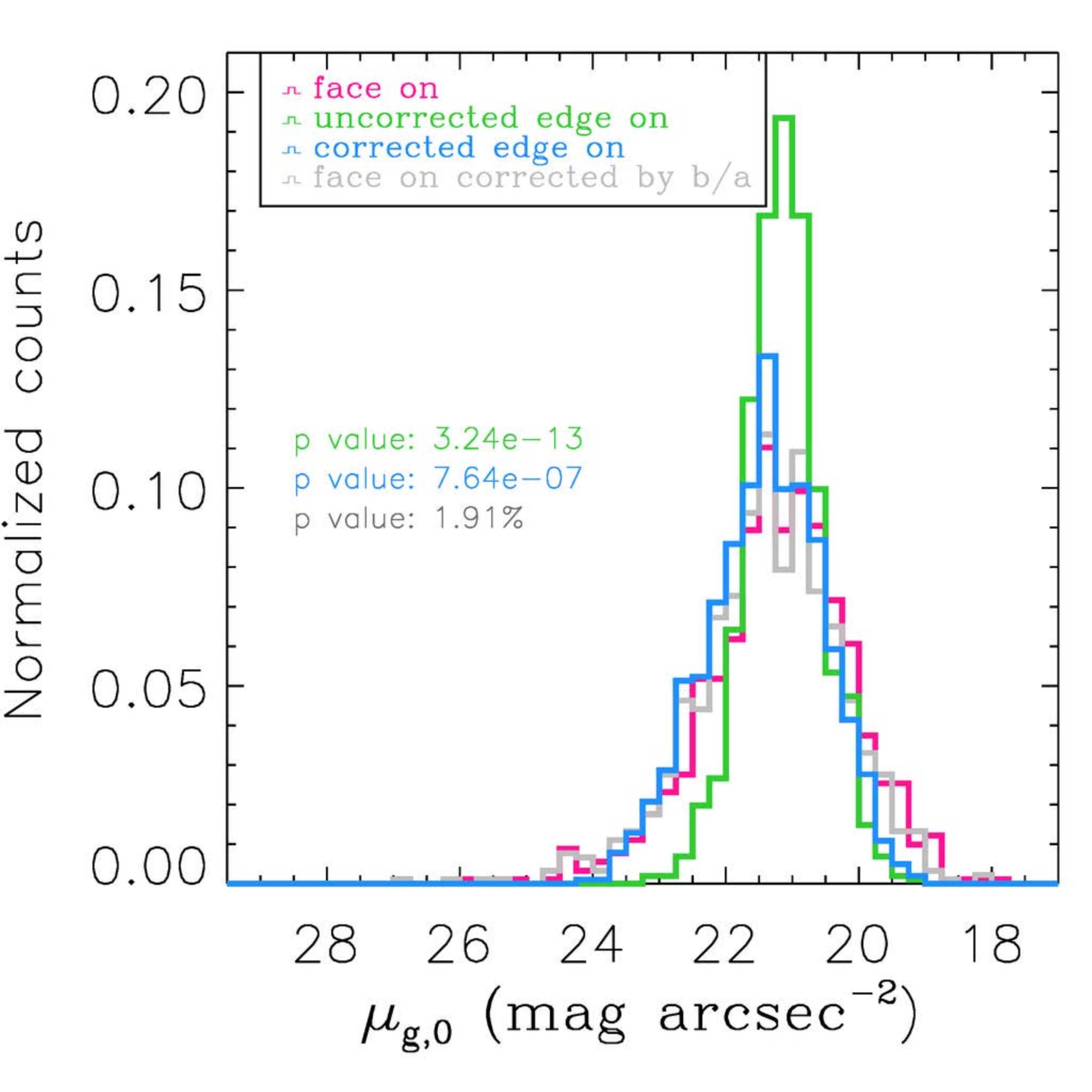}{0.33\textwidth}{(d)}
          \fig{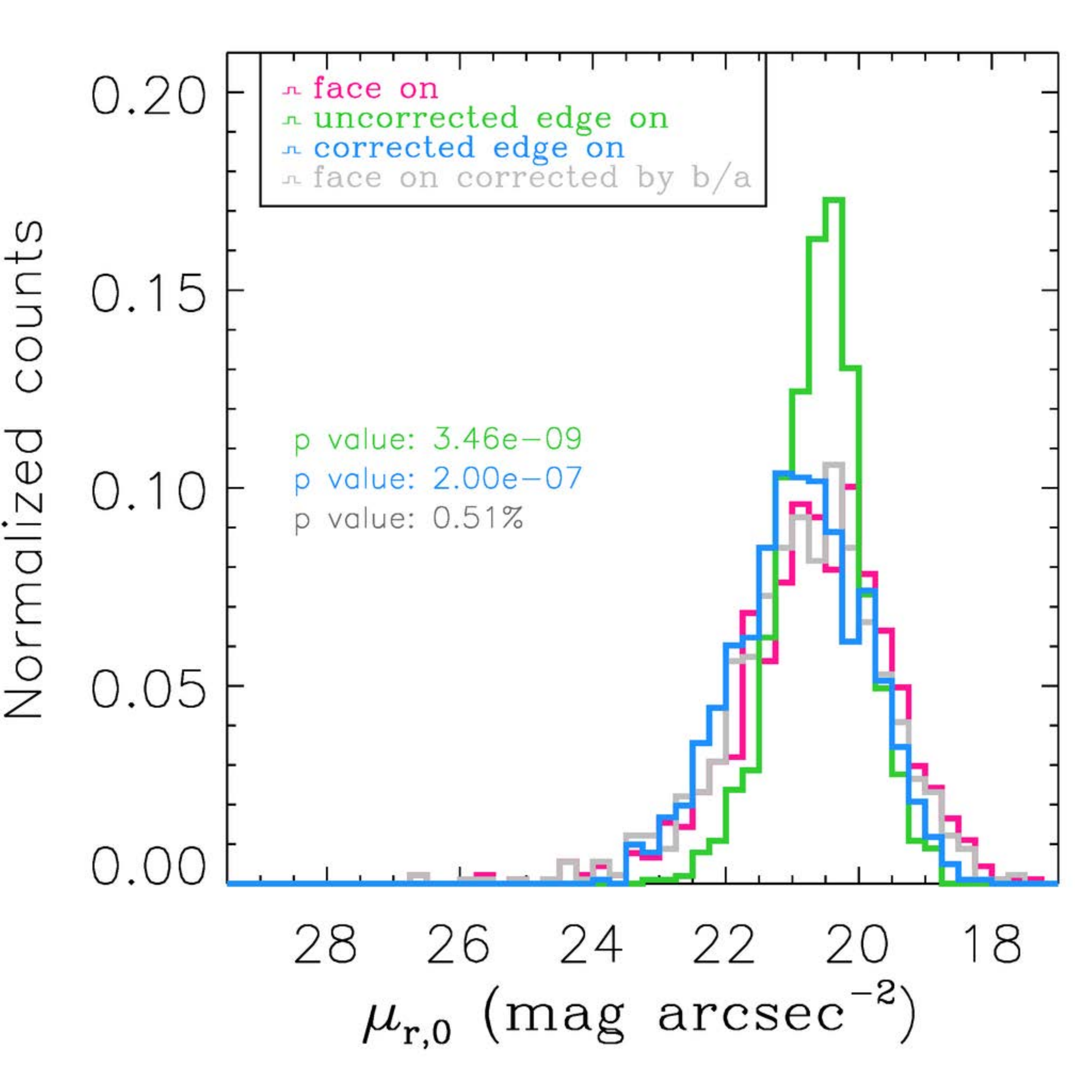}{0.33\textwidth}{(e)}
          \fig{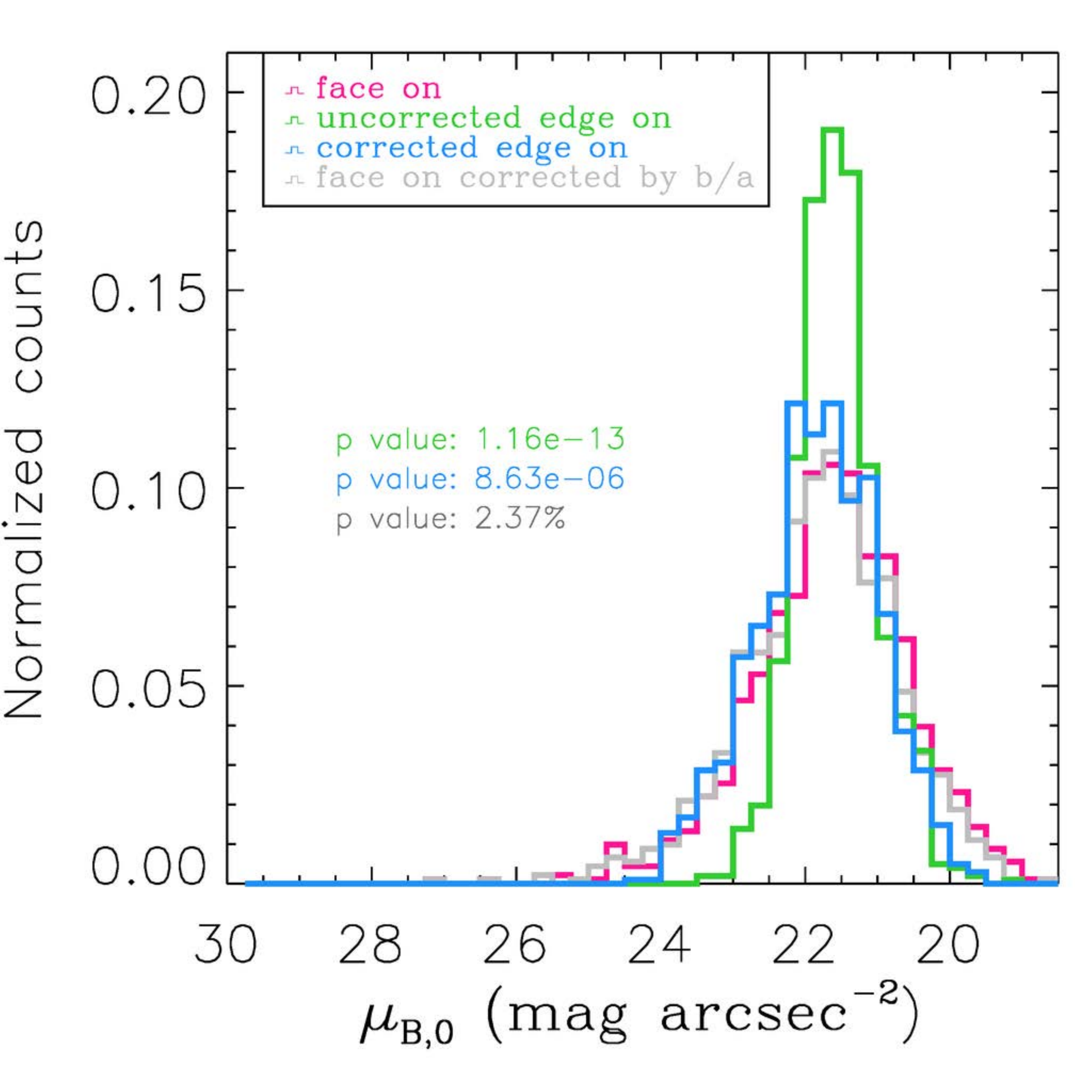}{0.33\textwidth}{(f)}
         }
\caption{Histograms of absolute magnitude (top panels) and central surface brightness (bottom panels) of face-on galaxies(red) and edge-on galaxies before(green)/after(blue) correction in $g/r/B$-band. The $p$-values are the K-S test results between face-on galaxies and edge-on galaxies before correction (green) and after correction (blue). The gray lines are the face-on central surface brightness that has been roughly corrected by $-2.5log_{10}(b/a)$ and the gray $p$-values are the test results between them and the corrected edge-on central surface brightness. From these panels, it can be seen that after correction, the differences between edge-on and face-on galaxies become much less than before.
\label{fig:mu_correct_hist}}
\end{center}
\end{figure*}

As shown by the black pluses in Figure \ref{fig:fit-result}(a-b), for both $g$ band and $r$ band, the $\Delta M$ (edge-on absolute magnitude minus face-on absolute magnitude) has a significant dependence on face-on absolute magnitudes, while the relationship between the $\Delta M$ and edge-on absolute magnitudes is not very obvious. For simplifying and clarifying physics, we use Equation \ref{eq:mag-correct} to fit for $\Delta M$.
\begin{equation}
\Delta M = a_{1} M_{\rm face}+\rm c_{1}\label{eq:mag-correct}
\end{equation}
The relationship between $\Delta (g-r)$ color and edge-on $g-r$ color is also clear, thus we build Equation \ref{eq:color-correct} to describe it,
\begin{equation}
\Delta (g-r) = a_{2} (g-r)_{\rm edge}+\rm c_{2}\label{eq:color-correct}
\end{equation}\\

There is also a relationship between $\Delta log_{10}(r_{s})$ and $log_{10}(r_{s,edge})$ as Figure \ref{fig:fit-result}(d-e) shown. Some previous literature suggested that the scale length of galaxy would also be affected by inclination and internal extinction \citep{Giovanelli1995, Lu1998, Padilla2008}. We have both tried the linear function of one independent variable, $log_{10}(r_{s,edge})$, and two independent variables, $log_{10}(r_{s,edge})$ and absolute magnitude. There is only a difference of $0.002\sim 0.004$ of the residual standard errors between these two functions. And all of them show a strong correlation with $log_{10}(r_{s,edge})$. So we choose the simple formula (Equation \ref{eq:lgrs-correct}) for the correction.
\begin{equation}
\Delta \log_{10}(r_{\rm s}) = a_{3} \log_{10}(r_{\rm s,edge})+\rm c_{3}.\label{eq:lgrs-correct}
\end{equation}\\

The rlm function of R, a robust linear function with MM-estimation, is chosen to fit these relationships as Equation \ref{eq:mag-correct} -- \ref{eq:lgrs-correct} represent. The best-fit results of these functions are shown in Table \ref{tbl:fit-results} and plotted as black lines in Figure \ref{fig:fit-result}. For testing the self-consistent of the fitting results of $\Delta M$ in different bands, we also calculate the $\Delta M_{\rm r}$ from the fitting results of $g$-band and $g-r$ color, and compare it with the direct fitting result of $\Delta M_{\rm r}$. As Figure \ref{fig:fit-result}(b) displays, the calculating results drawn with red plus signs, are in good agreement with the fitting result drawn with black line. Thus we suggest that the relationships between $\Delta M$ and $M_{\rm face}$ for both $g$ and $r$ band are feasible. Note that the $\Delta M$ would become negative with much fainter magnitude, we does not correct these magnitudes because the $M_{face}$ could not be fainter than $M_{edge}$ normally.\\

\begin{table}{htb!}
\caption{Coefficients and deviations of the fit of Eq. \ref{eq:mag-correct}--\ref{eq:lgrs-correct} \label{tbl:fit-results}}
\centering
\begin{tabular}{ccc}
\hline
\hline
    & $a_{1}$          & $c_{1}$  \\
\hline            
$g$ & $-0.30 \pm 0.02$ & $-4.99 \pm 0.46$\\
\hline            
$r$ & $-0.27 \pm 0.03$ & $-4.60 \pm 0.50$\\
\hline            
\hline            
    & $a_{2}$         &  $c_{2}$  \\
\hline            
    & $0.86 \pm 0.04$ & $-0.32 \pm 0.02$ \\
\hline
\hline
    & $a_{3}$         &  $c_{3}$  \\
\hline
$g$ & $0.35 \pm 0.04$ &  $-0.21 \pm 0.03$\\
$r$ & $0.32 \pm 0.04$ &  $-0.18 \pm 0.03$\\
\hline
\hline
\end{tabular}
\end{table}

\section{Sample and Properties}
\label{sec:sample-prop}

\subsection{Sample}
\label{subsec:sample}

Equation \ref{eq:mu} is adopted to calculate the central surface brightness for real face-on galaxies. For our edge-on galaxies, their $\mu_{\rm 0,edge}$ are converted into face-on $\mu_{\rm 0,face}$ using Equation \ref{eq:correct_mu}. The $\mu_{\rm 0,edge}$ are obtained from GALFIT output results with considering the galactic extinction. Corrections of internal extinction, surface brightness profile model, scale length and cosmological dimming effect are considered in the order of Equation \ref{eq:correct_mu}.\\

\begin{gather}
\mu_{0} = m + 2.5 \log_{10}(2\pi r_{\rm s}^{2})\label{eq:mu}\\
\mu_{\rm 0,corr} = \mu_{\rm 0,edge}-\Delta M-2.5 \log_{10}(h_{\rm s}/r_{\rm s})-5 \Delta \log_{10}(r_{\rm s})\notag\\
-10 \log_{10}(1+z)\label{eq:correct_mu}
\end{gather}\\

However, the observed data are edge-on values for these edge-on galaxies, so that the $\mu_{\rm 0,corr}$ cannot be calculated by directly using Equations \ref{eq:mag-correct}. We convert the equation with fitting results to Equation \ref{eq:mag-g-correct} and \ref{eq:mag-r-correct}, the functions of $M_{\rm edge}$ for both $g$ and $r$ band:\\
\begin{gather}
\Delta M_{\rm g} = -0.44 M_{\rm edge,g}-7.16\label{eq:mag-g-correct}\\
\Delta M_{\rm r} = -0.38 M_{\rm edge,r}-6.35\label{eq:mag-r-correct}.
\end{gather}
Then, the $g$-band and $r$-band central surface brightness are calculated by Equation \ref{eq:correct_mu} and $B$-band magnitude and central surface brightness are calculated by Equation \ref{eq:B-band-cal} \citep{Smith2002}, and Equation \ref{eq:B-band-cal-mu} \citep{Zhong2008, Du2015}:\\
\begin{gather}
m_{\rm B} = m_{\rm g} + 0.47 (m_{\rm g}-m_{\rm r})+0.17 \label{eq:B-band-cal}\\
\mu_{\rm 0,B} = \mu_{\rm 0,g}+0.47 (\mu_{\rm 0,g}-\mu_{\rm 0,r})+0.17\label{eq:B-band-cal-mu}
\end{gather}\\

For checking the corrections, the distributions of absolute magnitude and central surface brightness before and after correction are presented in Figure \ref{fig:mu_correct_hist}. The $p$-values in these panels are the results of two-sided Kolmogorov-Smirnov (K-S) test. The green values are the results for uncorrected edge-on galaxies comparing with face-on galaxies, while the blue values are the results for corrected edge-on galaxies comparing with face-on galaxies. As can seen from the top panels, the $p$-values for the corrected edge-on absolute magnitudes in $g/r/B$ band are much larger than those for uncorrected edge-on absolute magnitudes. This indicates that the differences of absolute magnitudes between edge-on and face-on galaxies become much less after correction.\\

However, for the central surface brightness (Figure \ref{fig:mu_correct_hist}(d-f)), there exists an offset between the corrected edge-on (blue lines) and the face-on (red lines). We suppose that is because these face-on galaxies are not completely face-on, and there are still some inclinations. In the work of \citetalias{Du2015}, the $\mu_0$ are calculated by the equation of $\mu_0 = m + 2.5 log_{10}(2\pi r_{\rm s}^2 q)-10 \log_{10}(1+z)$, in which the $q$ is the $b/a$. For confirming our conjecture, a roughly correction for inclination, $-2.5log_{10}(b/a)$, is added to the face-on central surface brightness to convert them into the results of a completely face-on galaxy. The distributions of the corrected face-on central surface brightness in $g/r/B$ band are represented by the gray lines in Figure \ref{fig:mu_correct_hist}(d-f). And the $p$-values in gray color are the K-S test results for central surface brightness between corrected face-on and corrected edge-on, and they are larger than blue values.\\

Although these gray $p$-values are far smaller than the criterion of $5\%$ to reject the null hypothesis, but we notice that these gray $p$-values are sensitive for a small change in the $b/a$ of face-on galaxies. If we simply add an offset value calculated by $-2.5log_{10}(b/a)$ to those face-on central surface brightness and change the $b/a$ in the range of $0.8$ to $1.0$, the $p$-values could get a largest result as $33.8\%$, which is far larger than $5\%$. We also create two different random samples from the face-on galaxies sample 1000 times and test on the distributions of central surface brightness of two samples. The least $p$-values could be $10.1\%$, which is smaller than $33.8\%$. So the possibility that the distributions of corrected edge-on central surface brightness and face-on central surface brightness are similar could be larger than the possibility that the distributions of central surface brightness of two random face-on samples are similar. This suggests that the correction of central surface brightness is maybe feasible.\\

Except the K-S test, we also have used the K-sample Anderson-Darling (A-D) test for testing differences, which is another sensitive two-sample test. The $p$-values of the A-D test are smaller than K-S test, such as $3.61e-21$ (A-D) compared to $1.16e-13$ (K-S) for uncorrected $\mu_{\rm 0,B,edge}$, and $0.09\%$ compared to $2.37\%$ for corrected $\mu_{\rm 0,B,edge}$. But the A-D test results also suggests that the correction could significantly reduce the differences between edge-on and face-on galaxies.\\

\begin{figure*}[htb!]
\begin{center}
\gridline{\fig{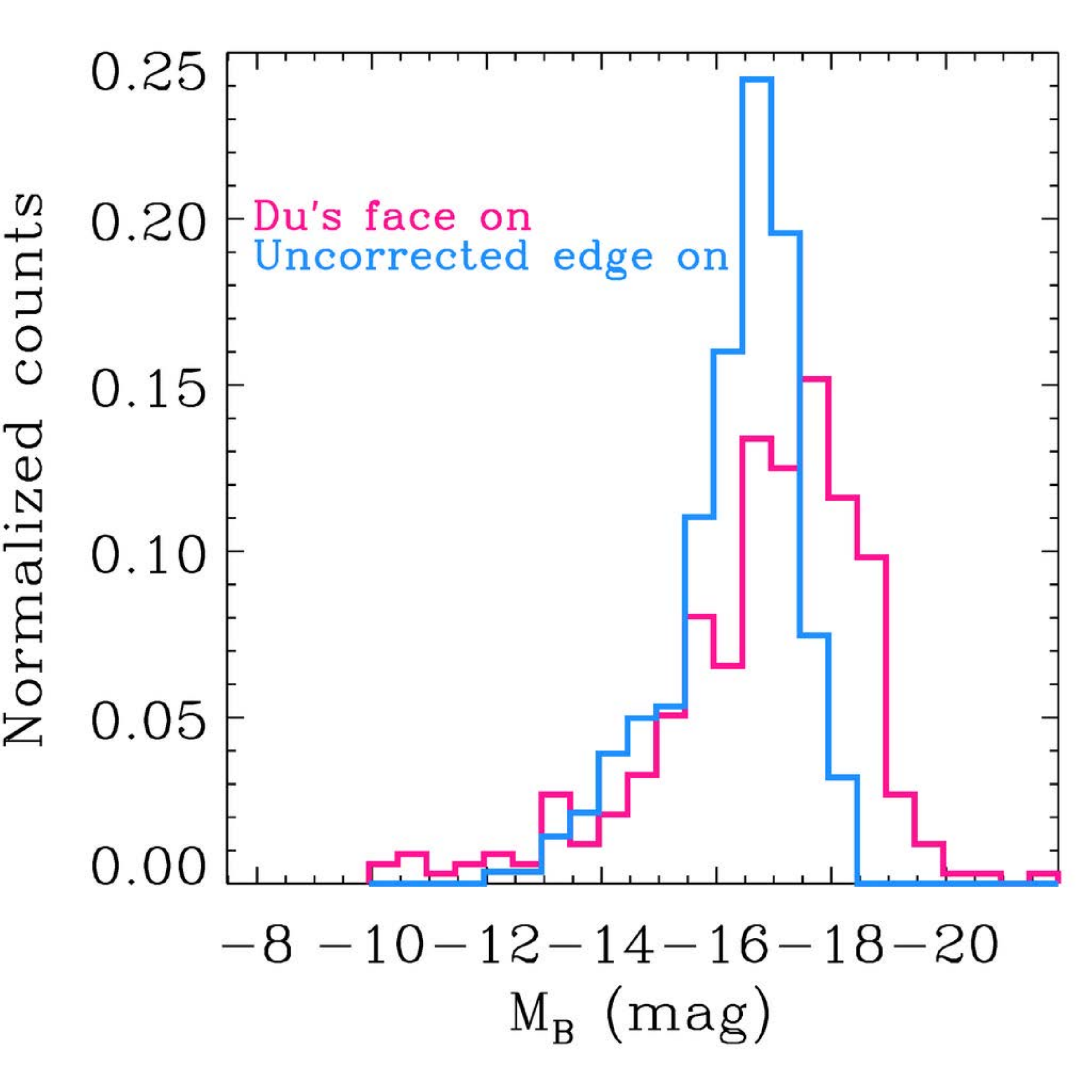}{0.25\textwidth}{(a)}
          \fig{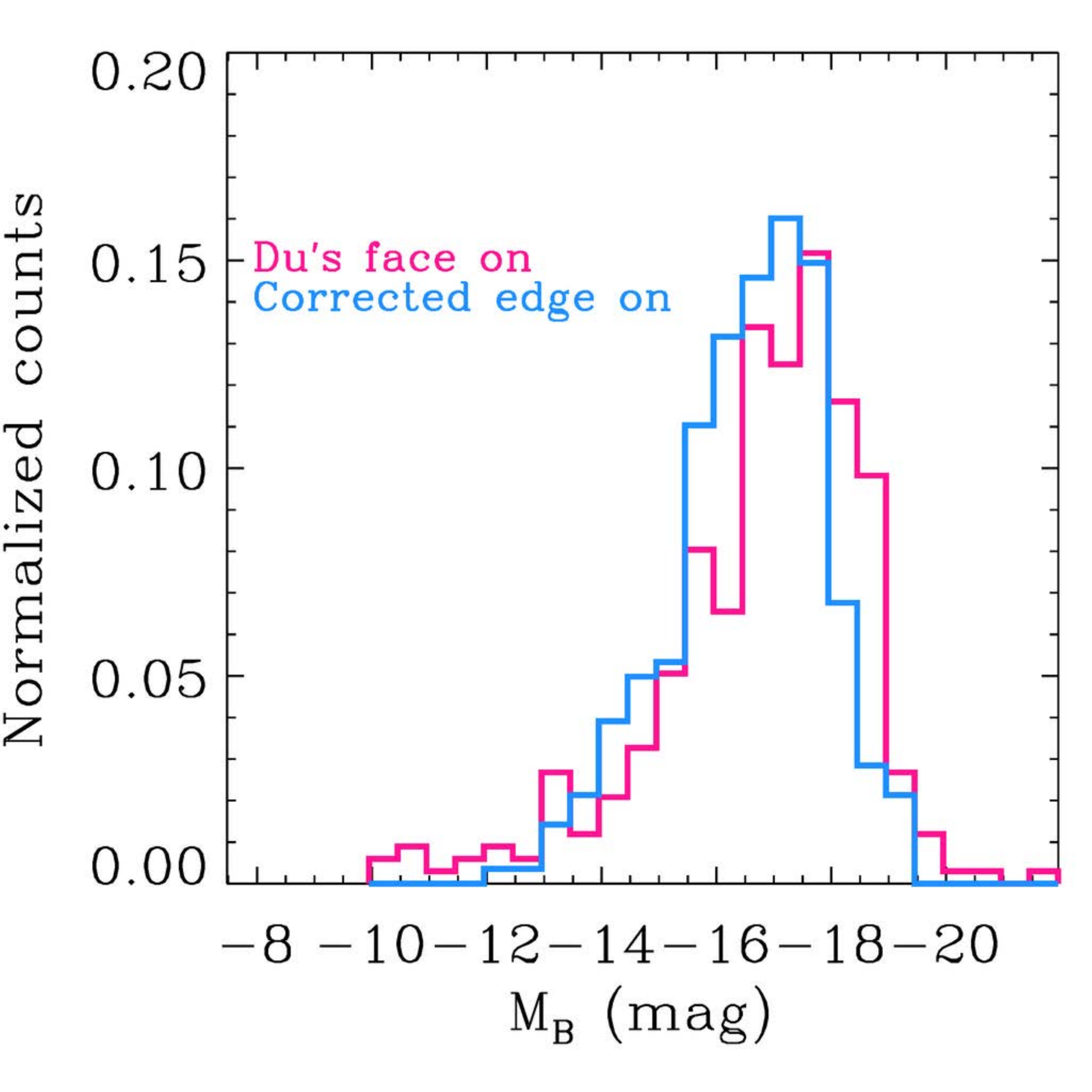}{0.25\textwidth}{(b)}
          \fig{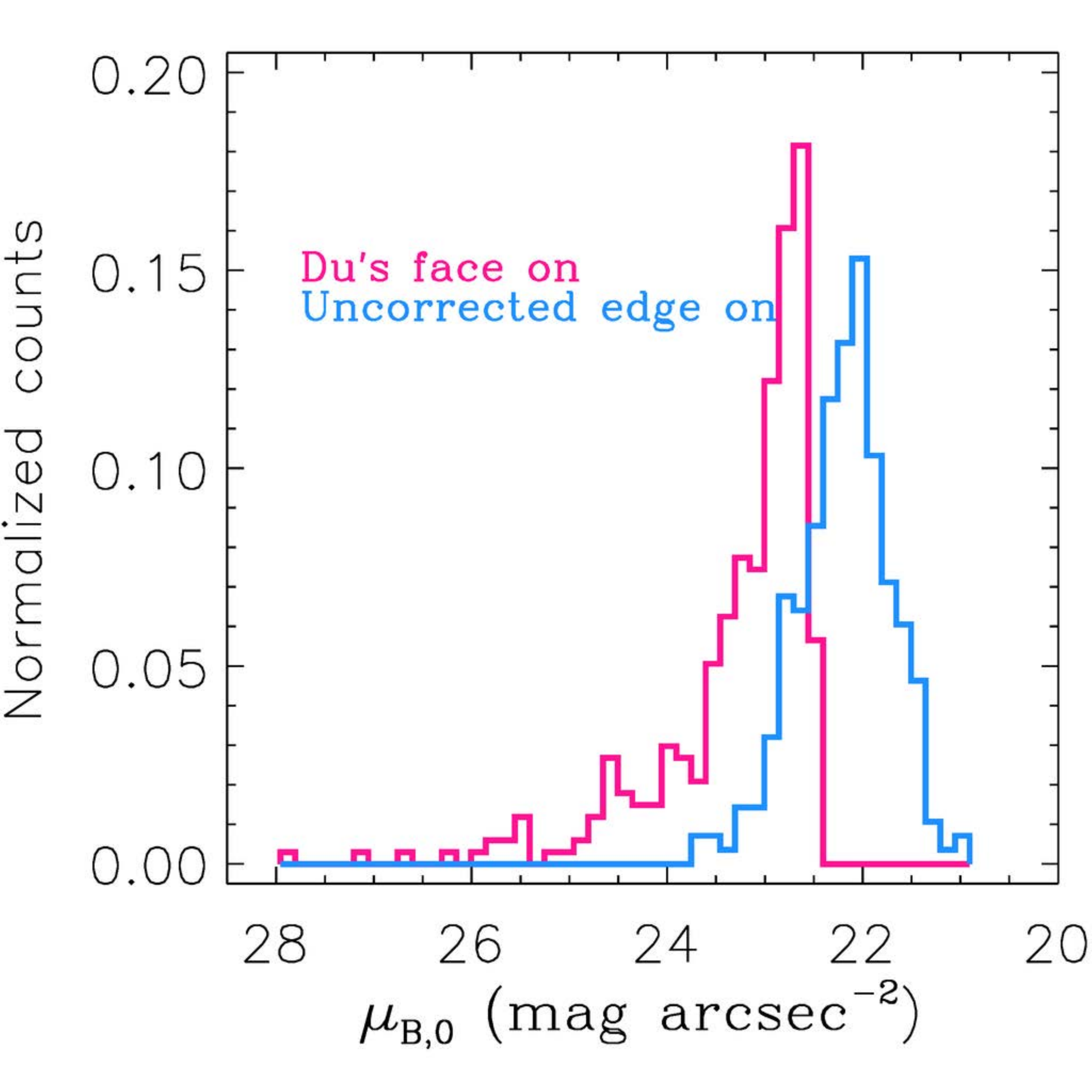}{0.25\textwidth}{(c)}
          \fig{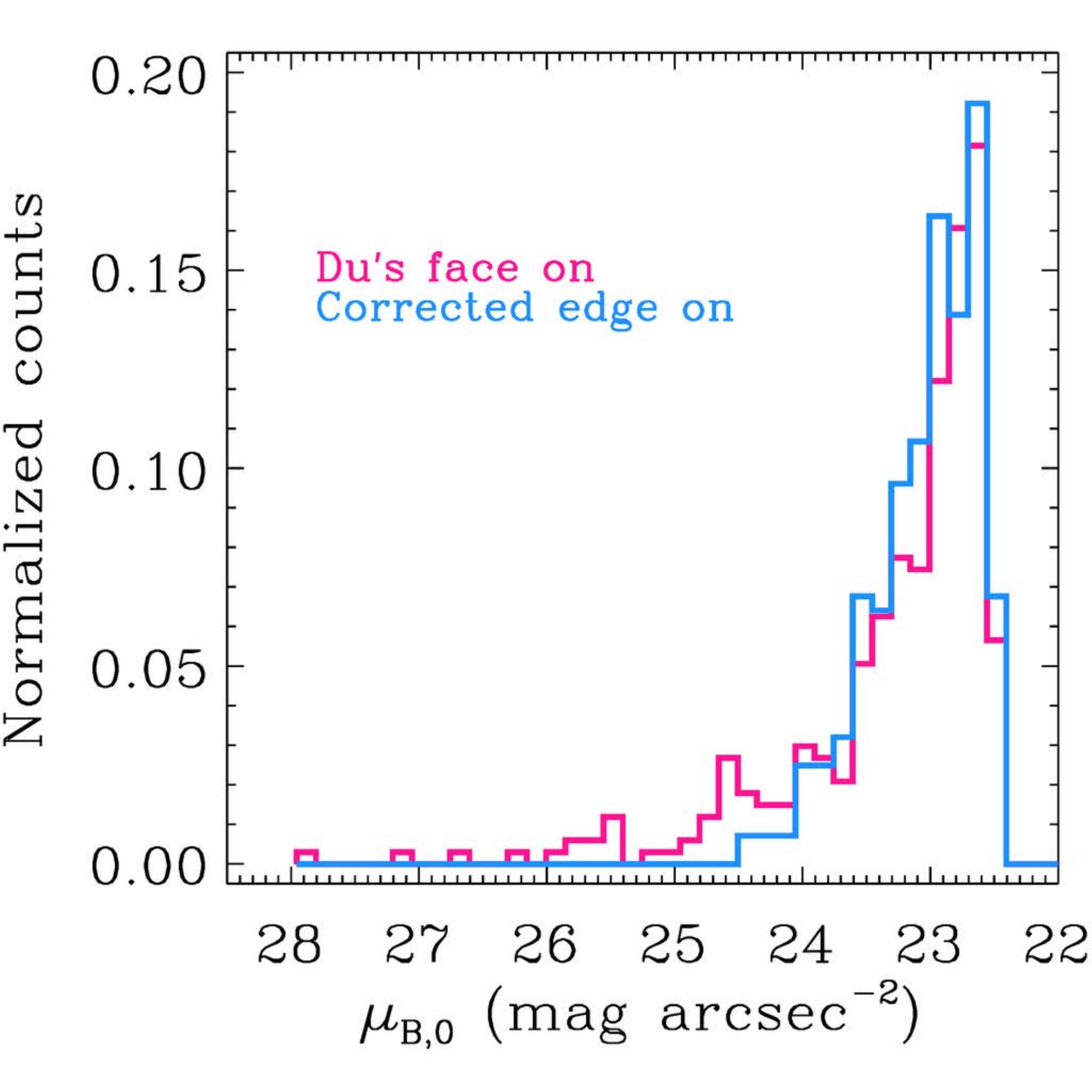}{0.25\textwidth}{(d)}
          }
\vspace{-0.5cm}
\gridline{\fig{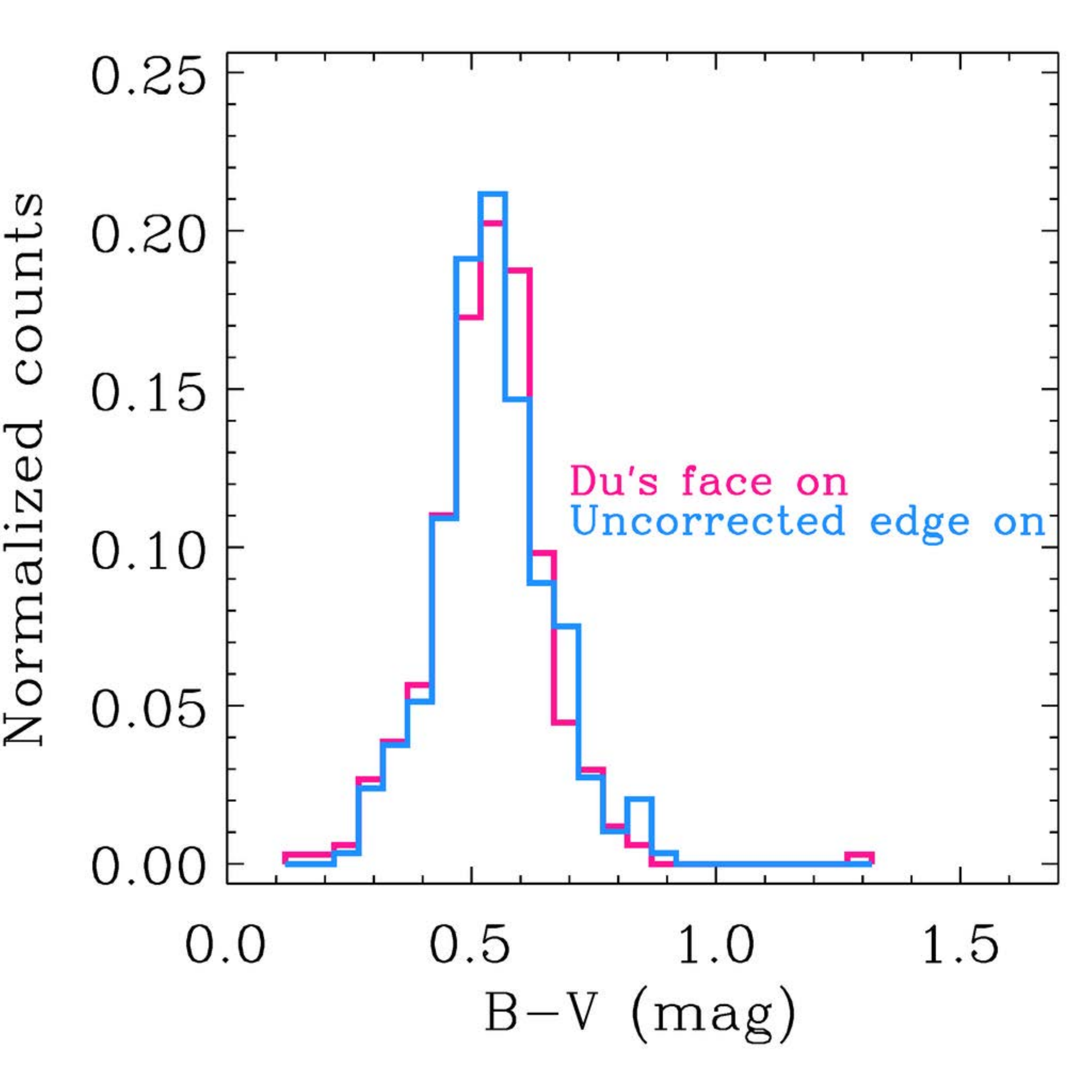}{0.25\textwidth}{(e)}
          \fig{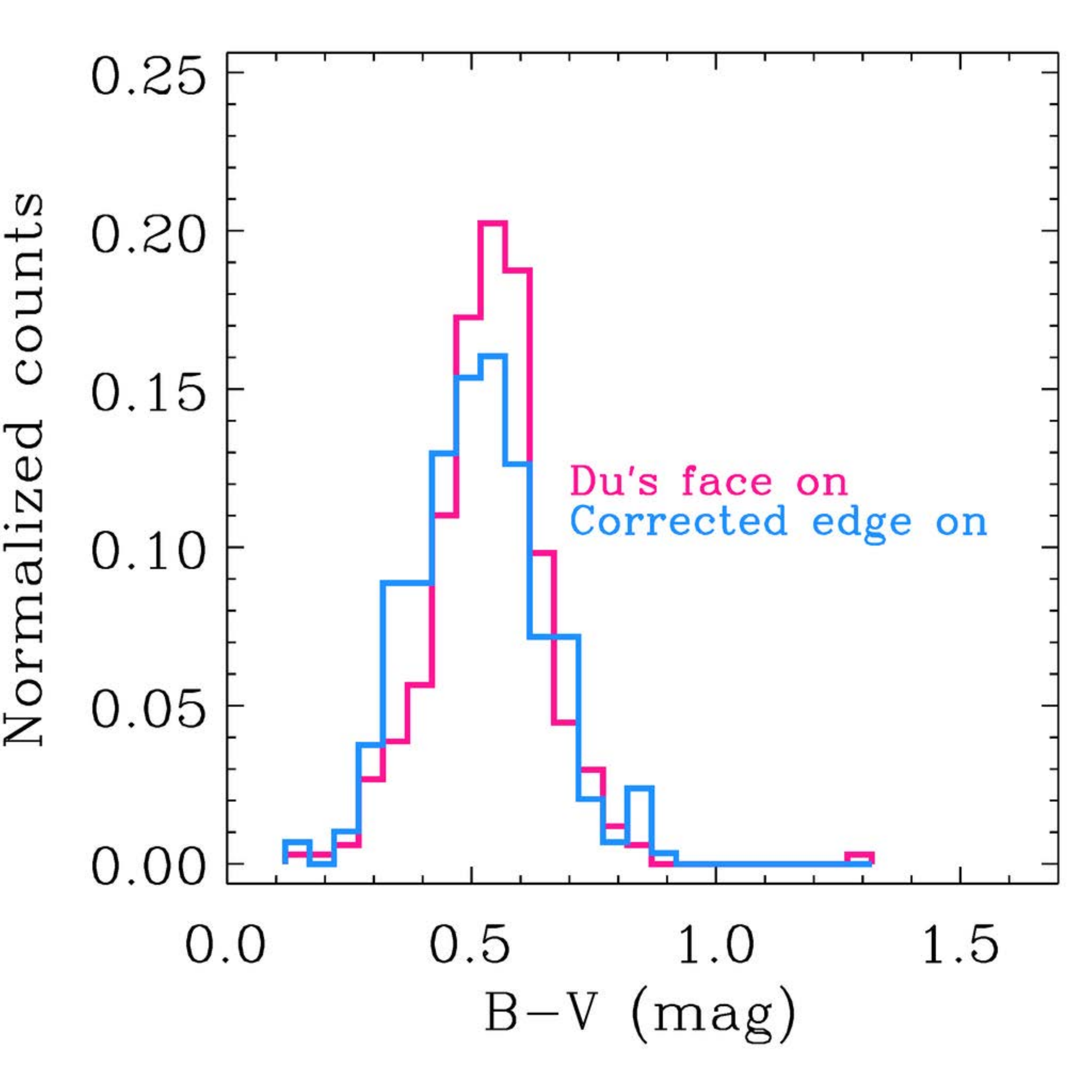}{0.25\textwidth}{(f)}
          \fig{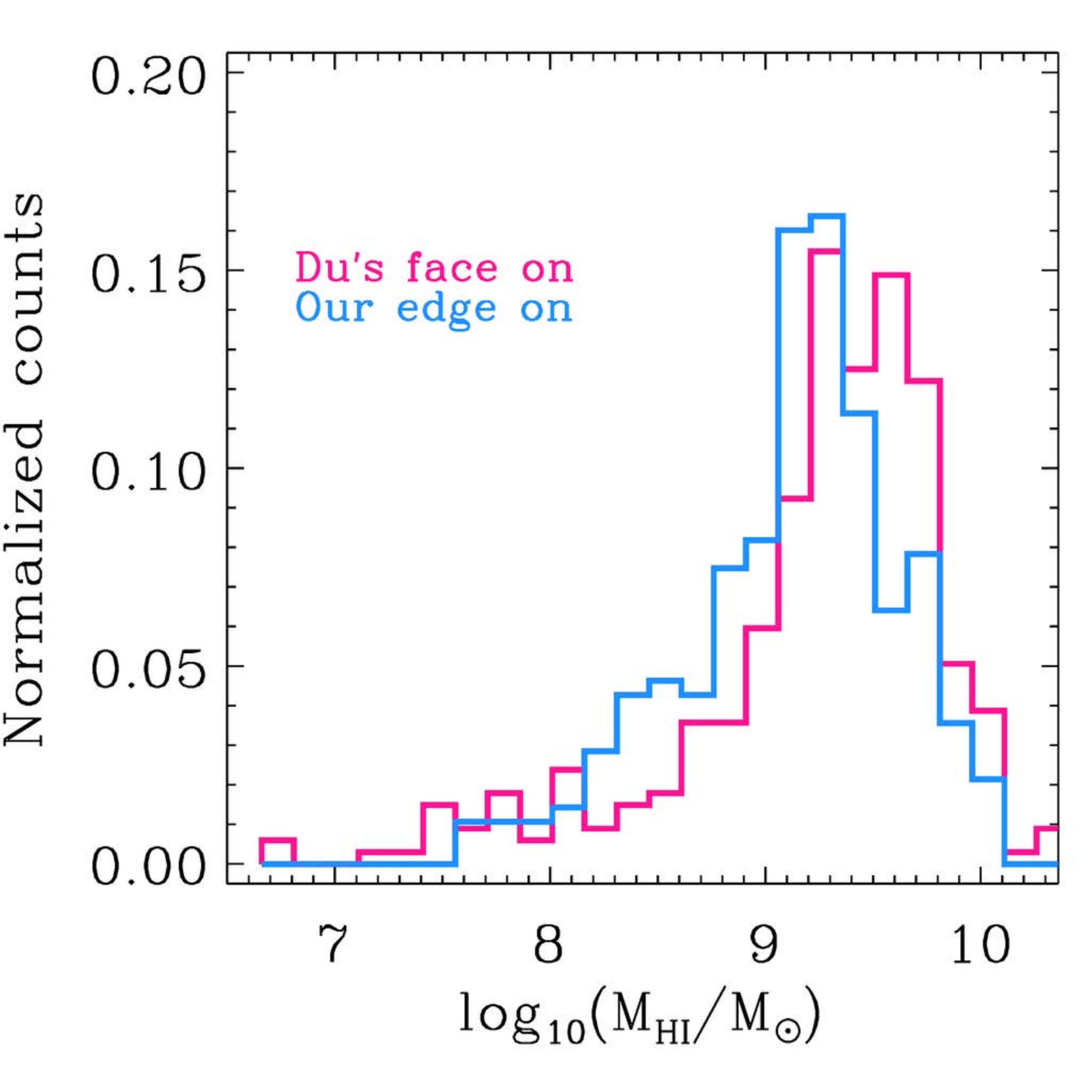}{0.25\textwidth}{(g)}
          \fig{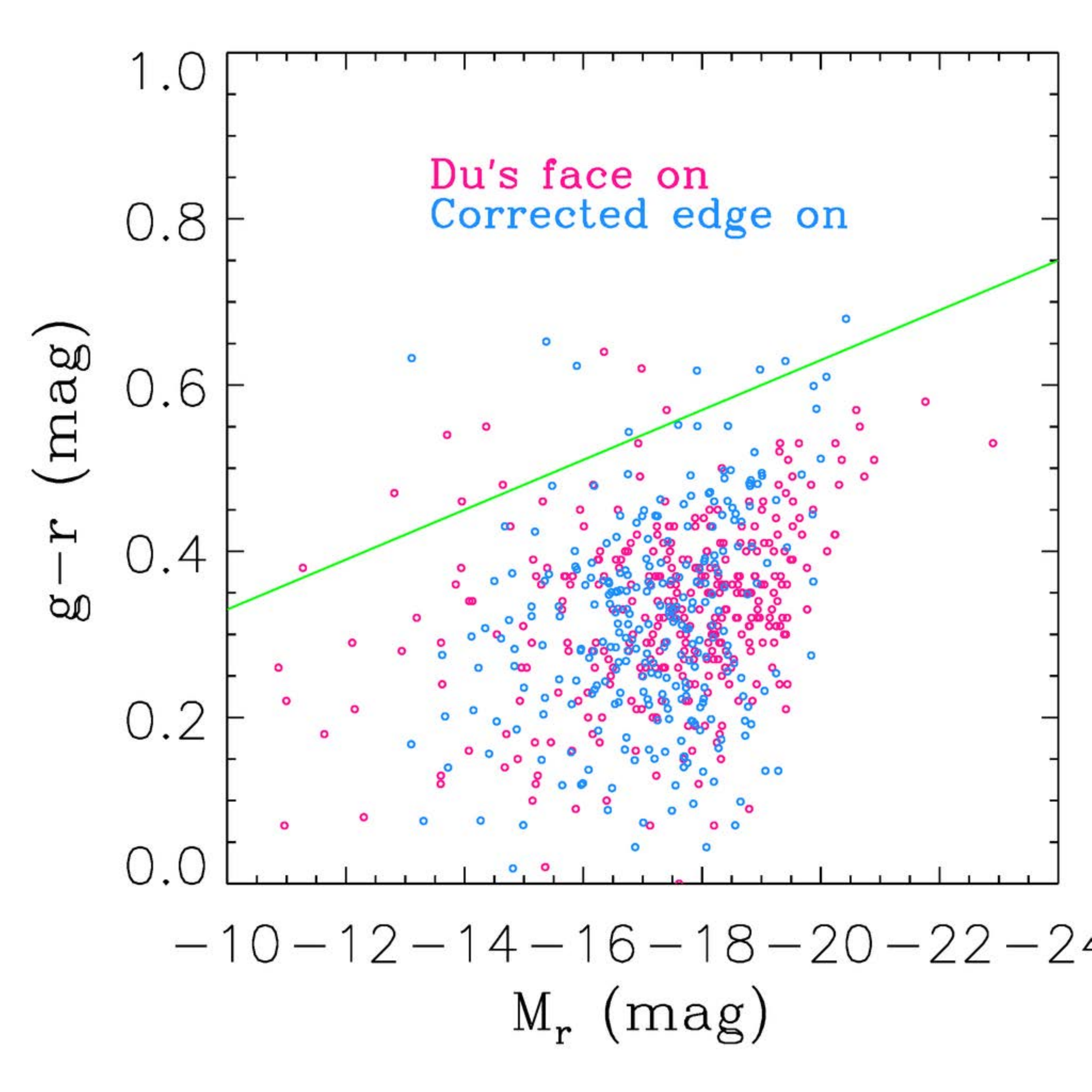}{0.25\textwidth}{(h)}
          }
\caption{(a)-(g): Comparison of the distributions of $B$-band absolute magnitude, $B$-band surface brightness and $B-V$ color of \citetalias{Du2015}'s face-on LSBGs and our edge-on LSBG candidates before and after correction; (h): The diagram of corrected $g-r$ vs. corrected $M_{\rm r}$. In these plots, blue lines and dots denote our edge-on LSBG candidates, red lines and dots denote \citetalias{Du2015}'s face-on LSBGs and the green line in panel (h) is the dividing line between ``red'' sequence galaxies and ``blue'' cloud galaxies \citep{Bernardi2010}. According to the panel (b), (d), (f) and (g), our edge-on LSBG candidates prefer to have bluer color and lower HI mass than \citetalias{Du2015}'s face-on LSBGs, but lack the faint galaxies with $\mu_{B,face} \geq 24.5\rm\ mag\cdot arcsec^{-2}$. The panel (h) indicates that most of our edge-on LSBG candidates are also located in the ``blue'' region.
\label{fig:properties}}
\end{center}
\end{figure*}

\begin{table*}[htb!]
\begin{rotatetable*}
\caption{The catalog of edge-on H{\sc{i}}-rich LSBG candidates sample \label{tbl:subset}}
\setlength{\leftskip}{-150pt}
\resizebox{\textheight}{!}{
\centering
\begin{tabular}{lllllllllllllllllll}
\hline
\hline
 AGCNr 
 & R.A. & Dec. 
 & $cz$ 
 & $W_{50}$ 
 & $\log_{10}(M_{\rm HI}/M_{\odot})$ 
 & Dist\tablenotemark{a}
 & $m_{\rm g}$\tablenotemark{b}
 & $M_{\rm B}$\tablenotemark{c}
 & $M_{\rm B,corr}$\tablenotemark{d}
 & $B-V$\tablenotemark{e}
 & $\mu_{\rm 0,B}$\tablenotemark{f}
 & $\mu_{\rm 0,B,model}$\tablenotemark{g}
 & $\mu_{\rm 0,B,mag}$\tablenotemark{h}
 & $\mu_{\rm 0,B,corr}$\tablenotemark{i}
 & $h_{\rm s,g}$\tablenotemark{j}
 & $r_{\rm s,g}$\tablenotemark{k} 
 & $r_{\rm s,g,corr}$\tablenotemark{l} 
 & Class\tablenotemark{m}
 \\
 & (J2000) & (J2000)
 & (km$\cdot$s$^{-1}$)
 & (km$\cdot$s$^{-1}$)
 & 
 & (Mpc)
 & (mag) 
 & (mag) 
 & (mag) 
 & (mag)
 & (mag$\cdot$arcsec$^{-2}$) 
 & (mag$\cdot$arcsec$^{-2}$) 
 & (mag$\cdot$arcsec$^{-2}$) 
 & (mag$\cdot$arcsec$^{-2}$) 
 & (kpc) 
 & (kpc) 
 & (kpc)
 & \\
\hline
(1)&(2)&(3)&(4)&(5)&(6)&(7)&(8)&(9)&(10)&(11)&(12)&(13)&(14)&(15)&(16)&(17)&(18)&(19)\\
\hline
$   290$ & $00:29:08$ & $+15:53:56$ & $   764$ & $    82 \pm      2$ & $  8.50 \pm   0.17$ & $  12.8 \pm    2.5$ & $ 15.77 \pm   0.01$ & $-14.48 \pm   0.65$ & $-14.48 \pm   0.65(  0.65)$ & $  0.44 \pm   0.01(  0.01)$ & $ 22.65 \pm   0.01$ & $ 24.03 \pm   0.02$ & $ 24.03 \pm   0.02(  0.02)$ & $ 23.01 \pm   0.02(  0.21)$ & $  0.27 \pm   0.05$ & $  0.94 \pm   0.18$ & $  0.59 \pm   0.18(  0.19)$ & $D$ \\
$  2052$ & $02:34:26$ & $+25:16:10$ & $  5700$ & $   173 \pm      3$ & $  9.27 \pm   0.06$ & $  79.0 \pm    2.2$ & $ 16.85 \pm   0.01$ & $-17.34 \pm   0.10$ & $-17.90 \pm   0.10(  0.11)$ & $  0.38 \pm   0.02(  0.03)$ & $ 22.72 \pm   0.01$ & $ 23.69 \pm   0.03$ & $ 23.13 \pm   0.03(  0.06)$ & $ 22.57 \pm   0.03(  0.13)$ & $  1.28 \pm   0.04$ & $  3.10 \pm   0.09$ & $  2.43 \pm   0.09(  0.15)$ & $M$ \\
$  2221$ & $02:44:58$ & $+30:22:40$ & $   833$ & $   103 \pm      2$ & $  8.13 \pm   0.31$ & $  12.3 \pm    4.3$ & $ 16.02 \pm   0.01$ & $-14.11 \pm   1.17$ & $-14.11 \pm   1.17(  1.17)$ & $  0.52 \pm   0.01(  0.01)$ & $ 21.84 \pm   0.01$ & $ 23.35 \pm   0.02$ & $ 23.35 \pm   0.02(  0.02)$ & $ 22.65 \pm   0.02(  0.15)$ & $  0.16 \pm   0.06$ & $  0.59 \pm   0.21$ & $  0.43 \pm   0.21(  0.21)$ & $D$ \\
$  5504$ & $10:12:49$ & $+07:06:11$ & $  1545$ & $   147 \pm      3$ & $  8.80 \pm   0.09$ & $  24.7 \pm    2.3$ & $ 15.85 \pm   0.01$ & $-15.78 \pm   0.31$ & $-15.78 \pm   0.31(  0.33)$ & $  0.54 \pm   0.01(  0.02)$ & $ 22.21 \pm   0.01$ & $ 23.72 \pm   0.01$ & $ 23.72 \pm   0.01(  0.10)$ & $ 22.84 \pm   0.01(  0.21)$ & $  0.38 \pm   0.04$ & $  1.46 \pm   0.14$ & $  0.99 \pm   0.14(  0.15)$ & $D$ \\
$  5844$ & $10:43:56$ & $+28:08:49$ & $  1470$ & $   129 \pm      5$ & $  8.43 \pm   0.09$ & $  24.0 \pm    2.2$ & $ 15.64 \pm   0.00$ & $-15.86 \pm   0.31$ & $-15.86 \pm   0.31(  0.32)$ & $  0.70 \pm   0.01(  0.03)$ & $ 22.28 \pm   0.00$ & $ 23.61 \pm   0.01$ & $ 23.61 \pm   0.01(  0.10)$ & $ 22.70 \pm   0.01(  0.22)$ & $  0.46 \pm   0.04$ & $  1.54 \pm   0.14$ & $  1.02 \pm   0.14(  0.16)$ & $D$ \\
$  6862$ & $11:53:15$ & $+11:38:01$ & $  2732$ & $   187 \pm      3$ & $  9.39 \pm   0.19$ & $  56.2 \pm   11.7$ & $ 15.99 \pm   0.01$ & $-17.36 \pm   0.70$ & $-17.94 \pm   0.70(  0.70)$ & $  0.69 \pm   0.01(  0.02)$ & $ 22.31 \pm   0.01$ & $ 24.10 \pm   0.01$ & $ 23.52 \pm   0.01(  0.06)$ & $ 22.55 \pm   0.01(  0.21)$ & $  0.76 \pm   0.16$ & $  3.77 \pm   0.79$ & $  2.46 \pm   0.79(  0.81)$ & $M$ \\
$  7421$ & $12:21:56$ & $+11:58:02$ & $   152$ & $   106 \pm      9$ & $  8.43 \pm   0.08$ & $  16.7 \pm    1.2$ & $ 15.60 \pm   0.01$ & $-15.17 \pm   0.24$ & $-15.17 \pm   0.24(  0.25)$ & $  0.59 \pm   0.01(  0.02)$ & $ 23.48 \pm   0.01$ & $ 24.24 \pm   0.02$ & $ 24.24 \pm   0.02(  0.05)$ & $ 23.10 \pm   0.02(  0.25)$ & $  0.73 \pm   0.05$ & $  1.44 \pm   0.10$ & $  0.86 \pm   0.10(  0.13)$ & $D$ \\
$  8575$ & $13:35:45$ & $+08:58:06$ & $  1163$ & $   120 \pm      3$ & $  8.89 \pm   0.26$ & $  15.2 \pm    4.5$ & $ 14.64 \pm   0.00$ & $-15.95 \pm   0.99$ & $-15.95 \pm   0.99(  1.00)$ & $  0.52 \pm   0.01(  0.02)$ & $ 22.54 \pm   0.00$ & $ 24.11 \pm   0.01$ & $ 24.11 \pm   0.01(  0.10)$ & $ 22.63 \pm   0.01(  0.34)$ & $  0.48 \pm   0.14$ & $  1.84 \pm   0.54$ & $  0.98 \pm   0.54(  0.55)$ & $D$ \\
$  9902$ & $15:34:33$ & $+15:07:59$ & $  1694$ & $   109 \pm      2$ & $  9.03 \pm   0.12$ & $  32.1 \pm    4.1$ & $ 16.58 \pm   0.01$ & $-15.68 \pm   0.43$ & $-15.68 \pm   0.43(  0.44)$ & $  0.44 \pm   0.02(  0.03)$ & $ 22.92 \pm   0.01$ & $ 24.03 \pm   0.02$ & $ 24.03 \pm   0.02(  0.10)$ & $ 23.29 \pm   0.02(  0.19)$ & $  0.60 \pm   0.08$ & $  1.57 \pm   0.20$ & $  1.14 \pm   0.20(  0.21)$ & $D$ \\
$100726$ & $00:13:39$ & $+15:40:29$ & $  1944$ & $   119 \pm      3$ & $  8.79 \pm   0.09$ & $  27.3 \pm    2.3$ & $ 16.51 \pm   0.01$ & $-15.37 \pm   0.28$ & $-15.37 \pm   0.28(  0.29)$ & $  0.49 \pm   0.02(  0.02)$ & $ 22.48 \pm   0.01$ & $ 23.94 \pm   0.03$ & $ 23.94 \pm   0.03(  0.07)$ & $ 23.28 \pm   0.03(  0.16)$ & $  0.34 \pm   0.03$ & $  1.21 \pm   0.10$ & $  0.91 \pm   0.10(  0.11)$ & $D$ \\
\hline
\hline
\end{tabular}}
\vspace{-0.2cm}
\begin{flushleft}
\textbf{Notes}.\\
The errors of $m_{\rm g/r}$, $\mu_{\rm 0,g/r}$ and $r_{\rm s,g}$ are obtained from SExtractor and GALFIT, and the errors of H{\sc{i}} mass, Distance and $w_{50}$ are provided from $\alpha .40$ catalog. And these errors propagate by propagation functions to obtain the other errors outside the brackets. The values in brackets are the final errors with adding the uncertainty produced by Monte-Carlo simulation in Section \ref{subsec:error}. 
\tablenotetext{a}{Dist, $W_{50}$ and H{\sc{i}} mass and their errors are achieved directly from the $\alpha. 40$ catalog.}
\tablenotetext{b}{$g$-band apparent magnitude which is measured directly from the photometry by using SExtractor.}
\tablenotetext{c}{$B$-band absolute magnitude before correction which is calculated from $M_{\rm g}$ and $M_{\rm r}$.}
\tablenotetext{d}{$B$-band absolute magnitude after correcting the internal extinction effect.}
\tablenotetext{e}{$B-V$ color with correction of extinction.}
\tablenotetext{f}{$B$-band observational edge-on central surface brightness which is calculated from the $\mu_{\rm 0,g}$ and $\mu_{\rm 0,r}$, and has been corrected the cosmological dimming effect and galactic extinction.}
\tablenotetext{g}{$B$-band central surface brightness corrected from observational edge-on values to face-on perspective. It is only considered the model correction of $\mu_{0}$.}
\tablenotetext{h}{$B$-band central surface brightness corrected from observational edge-on values to face-on perspective. It is considered both model correction of $\mu_{0}$ and the internal extinction correction.}
\tablenotetext{i}{$B$-band central surface brightness corrected from observational edge-on values to face-on perspective. It is considered all the corrections: model correction of $\mu_{0}$, internal extinction correction and scale length correction.}
\tablenotetext{j}{$g$-band scale height of galaxies in the unit of kpc obtained from GALFIT fitting with edge-on disk model.}
\tablenotetext{k}{$g$-band scale length of galaxies in the unit of kpc obtained from GALFIT fitting with edge-on disk model.}
\tablenotetext{l}{$g$-band scale length after correction.}
\tablenotetext{m}{we classify our edge-on LSBGs into dwarf galaxies(`D')(in which UDGs are also marked as `U'), giant LSBGs(`G') and moderate-luminosity galaxies(`M').}
\end{flushleft}
\end{rotatetable*}
\end{table*}

Then, follow the work of \citetalias{Du2015}, we select LSBGs by the criterion of $\mu_{\rm 0,B,corr} \geq 22.5$ mag$\cdot$arcsec$^{-2}$. Finally, 281	 LSBG candidates are selected. The basic information of these edge-on LSBG candidates are listed in Table 2, which includes the coordinates, HI parameters, optical photometry results and our corrected results.\\

To compare our edge-on sample with \citetalias{Du2015}'s work, we have selected 336 face-on LSBGs from \citetalias{Du2015}'s non-edge-on LSBGs sample by $b/a \geq 0.8$ \citep{Yoshino2015}. Plot the distributions of some optical and H{\sc{i}} properties in Figure \ref{fig:properties} for both edge-on and face-on LSBGs, and describe the comparison in following sections (4.2-4.4).

\subsection{Optical properties}
\label{subsec:opt-pro}

The $B$-band corrected absolute magnitude of our edge-on galaxies spans from $-19.26 \sim -12.01\rm\ mag$, with an average of $\langle M_{\rm B,corr}\rangle=-16.55$ mag. While the range of \citetalias{Du2015}'s face-on LSBGs sample, $-21.95\sim -9.98\rm\ mag$, is wider than our sample. From Figure \ref{fig:properties}, the central surface brightness of \citetalias{Du2015}'s face-on LSBGs also spans a wider range than our sample. The faintest $\mu_{\rm 0,B,corr}$ of our sample is only $24.46 \rm\ mag\cdot arcsec^{-2}$.\\

The $B-V$ of our edge-on galaxies, as Figure \ref{fig:properties}(f) shown, spans from $0.13$ to $0.89$ after correcting. And the mean value $\langle B-V \rangle=0.52$ is slightly bluer than that of \citetalias{Du2015}'s face-on galaxies, which is $0.54$. Here, the $B-V$ color is calculated by equation $B-V = 1.02(g-r) + 0.20$ \citep{Smith2002}.\\

\subsection{HI properties}
\label{subsec:hi-pro}

The H{\sc{i}} information from the ALFALFA catalog are provided on the ALFALFA website\footnote{\url{http://egg.astro.cornell.edu/alfalfa/data/index.php}}. The H{\sc{i}} mass of our edge-on LSBGs are in the range of $7.64$ to $10.11$ dex, with an average of $9.12$. There are more medium H{\sc{i}} mass galaxies ($7.7 < \log_{10}(M_{\rm HI}/M_{\odot}) < 9.5$)\citep{Huang2012} among our edge-on LSBG candidates, nearly 79\%(221/281) compared to 58\% of the face-on LSBGs selected by \citetalias{Du2015}. Except the medium H{\sc{i}} mass galaxies, there are 57 galaxies (20\%) with high mass of H{\sc{i}} gas ($\log_{10}(M_{\rm HI}/M_{\odot}) \geq 9.5$) and 3 (1\%) galaxies with low mass of H{\sc{i}} gas ($\log_{10}(M_{\rm HI}/M_{\odot}) \leq 7.7$), compared to 38\% and 4\% of the Du+15's face-on LSBGs respectively.\\

\subsection{Classification of the edge-on LSBGs}
\label{subsec:class}

Since LSBGs is a subclass of galaxies selected only based on central surface brightness. It indicates that LSBGs are a heterogeneous class of galaxies. All these edge-on LSBG candidates we selected have been classified, as also conducted in \citetalias{Du2015}, into dwarf LSBGs ($M_{\rm B} \geq -17$) \citep[][and references therein]{Sandage1984, Impey1988, Dunn2010}, giant LSBGs ($M_{\rm B} \leq -19$) \citep{Bothun1987, Bothun1990, Sprayberry1993, Sabatini2003, Galaz2015, Boissier2016}, and moderate-luminosity LSBGs ($-19 < M_{\rm B} < -17$). Finally, there are 167 (59.4\%) dwarf LSBGs, 6 (2.1\%) giant LSBGs and 108 (38.4\%) moderate-luminosity LSBGs have been selected, and the classification is listed in the column `Class' of Table \ref{tbl:subset}.\\

Compared with \citetalias{Du2015}'s face-on LSBGs, of which 46.1\% are dwarf galaxies, our edge-on LSBG candidates sample tends to have more dwarf galaxies. In addition to these three types, ultra-diffuse galaxies (UDGs) is a currently special population with very extended shape and very faint luminosity, which is included in dwarf galaxies \citep{van.Dokkum2015}. And we have already selected eleven edge-on UDG candidates from the $\alpha.40$ catalog in \citet{He2019}. These UDG candidates have been tabbed as `U' in Table \ref{tbl:subset}. While the Galactic extinction have not been corrected in \citet{He2019}, we have considered them in this work.\\

\section{Discussion}
\label{sec:discussion}

\subsection{Uncertainties of Sample}
\label{subsec:error}

Except the errors listed outside the brackets in Table \ref{tbl:subset}, which are generated by the photometric and profile fitting software (SExtractor and GALFIT) or from the ALFALFA catalog, there are also some other uncertainties in our sample.\\

An uncertainty brought by the profile model fitting may exist if the exponential model is not suitable for the brightness profile of a galaxy. Fortunately, our edge-on LSBG candidates are well fitted by the edge-on exponential model of GALFIT, 96\% of the sample have $\chi ^2$ below 1.4, and the maximum of $\chi ^2$ is 1.9. The residual images of all these galaxies have been well object-subtracted with visual checking. So the model fitting will not take a significant uncertainty to our sample.\\

The most influential uncertainty is caused by the internal extinction and scale length correction, and that is the reason why our edge-on LSBGs are called LSBG candidates. For estimating this uncertainty, a particular case of Monte-Carlo method, bootstrap, has been applied to the whole process from making the face-on-edge-on matched catalog to obtaining the corrected results for edge-on galaxies. We have selected 1013 edge-on points and 907 face-on points from these samples randomly and independently. The numbers of points are the same as the numbers of edge-on and face-on sample. Repeat the sampling 10,000 times to do the following work. Finally, we obtain the standard deviations of these 10,000 results as the uncertainty caused by the correction, add them to the above errors and list the final errors in the brackets in Table \ref{tbl:subset}. From the catalog, the final errors of the corrected $B$-band central surface brightness are in the range of $0.05\sim 0.34$, and 90\% are less than $0.16$.\\

Since the ratio of scale height to scale length of thick disk galaxies are relatively large, when they are on edge-on perspective, the $b/a$ will be larger than 0.3, which is beyond the criterion of $b/a \leq 0.3$. Such a criterion may lead to a missing of thick disk LSBGs.\\

\subsection{Fraction of LSBGs}
\label{subsec:lsbg-frac}

\begin{figure}[htb!]
\plotone{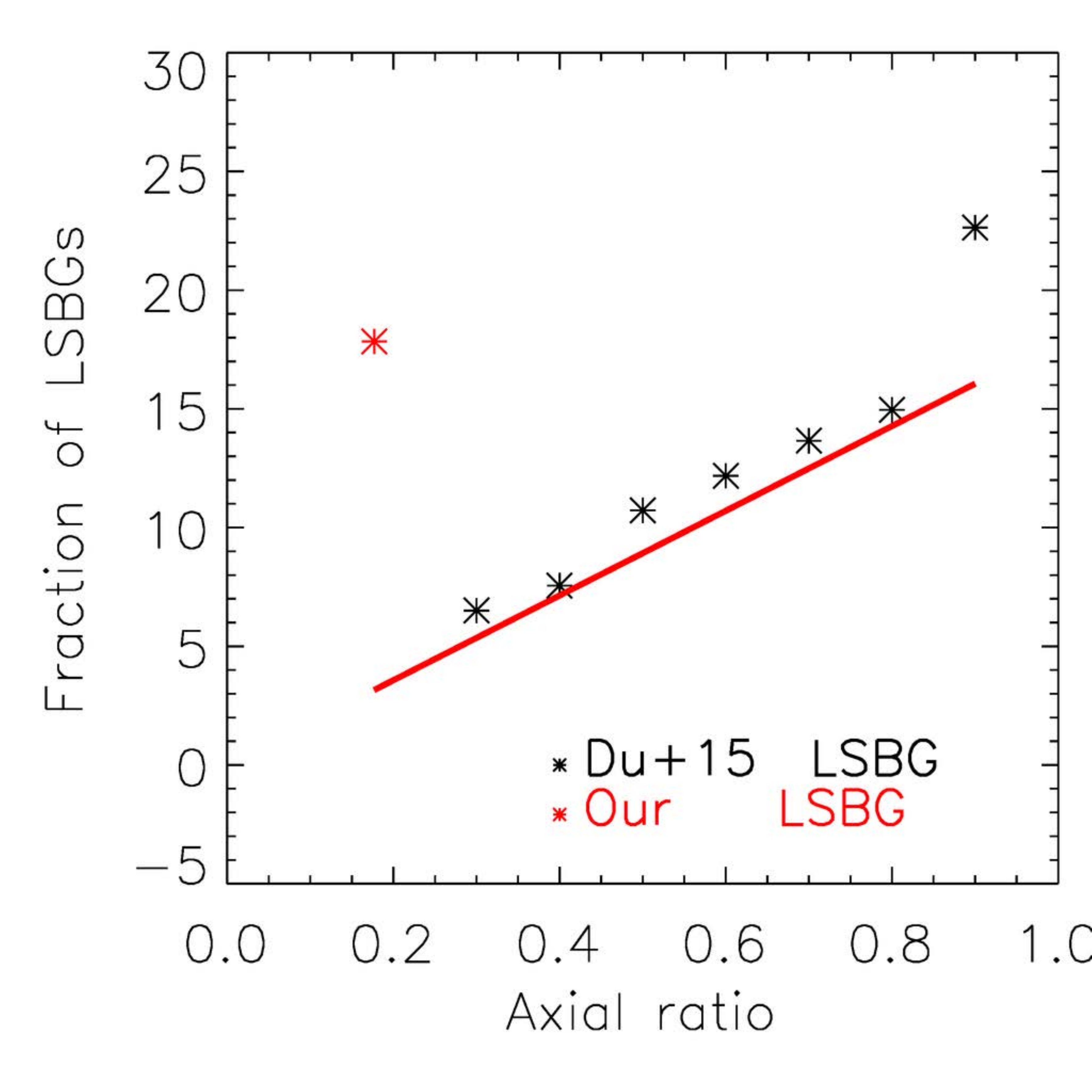}
\caption{This figure presents the correlation between fraction of LSBGs and axial ratio. Black points are fractions of \citetalias{Du2015}'s non-edge-on sample, which have not corrected the central surface brightness for galaxies which are not actually face-on galaxies. The red point is the fraction of our edge-on LSBG candidates (17.8\%). The red line presents the function of $17.8\% \times b/a$. It shows that there exists a tight correlation between LSBG fraction and axial ratio and suggests that without correction, galaxies with high axial ratio are biased to beyond the threshold of LSBGs. \label{fig:frac-lsbg}}
\end{figure}

There are 281 LSBG candidates selected from the 1575 edge-on galaxies sample, leading to a fraction of $\sim 17.8\%$. This is higher than the fraction of the non-edge-on LSBGs in \citetalias{Du2015} ($\sim 12\%$). However, we note that the inclination of the LSBGs in \citetalias{Du2015}'s non-edge-on sample spans a wide range. The fraction of observed LSBGs decreases with decreasing axial ratio $b/a$ and could be $22.6\%$ when their $b/a$ approach to $1$.\\

In Figure \ref{fig:frac-lsbg} shown, the red line, which presents the function $17.8\% \times b/a$, is consistent with the distribution of fractions of Du15's non-edge-on sample in different b/a bins. This suggests that there exists an obvious correlation between LSBG fraction and axial ratio, hence the decrease of the fraction of LSBGs may mostly because of the inclination. There were only 75 edge-on LSBGs would meet the selection criteria if these edge-on galaxies are not corrected into face-on perspective. Thus, if the inclination effects are not considered, a large fraction of LSBGs will be lost.\\

\subsection{Compare with HSBGs}
\label{subsec:L-HSBG}

\begin{figure*}[htb!]
\gridline{\fig{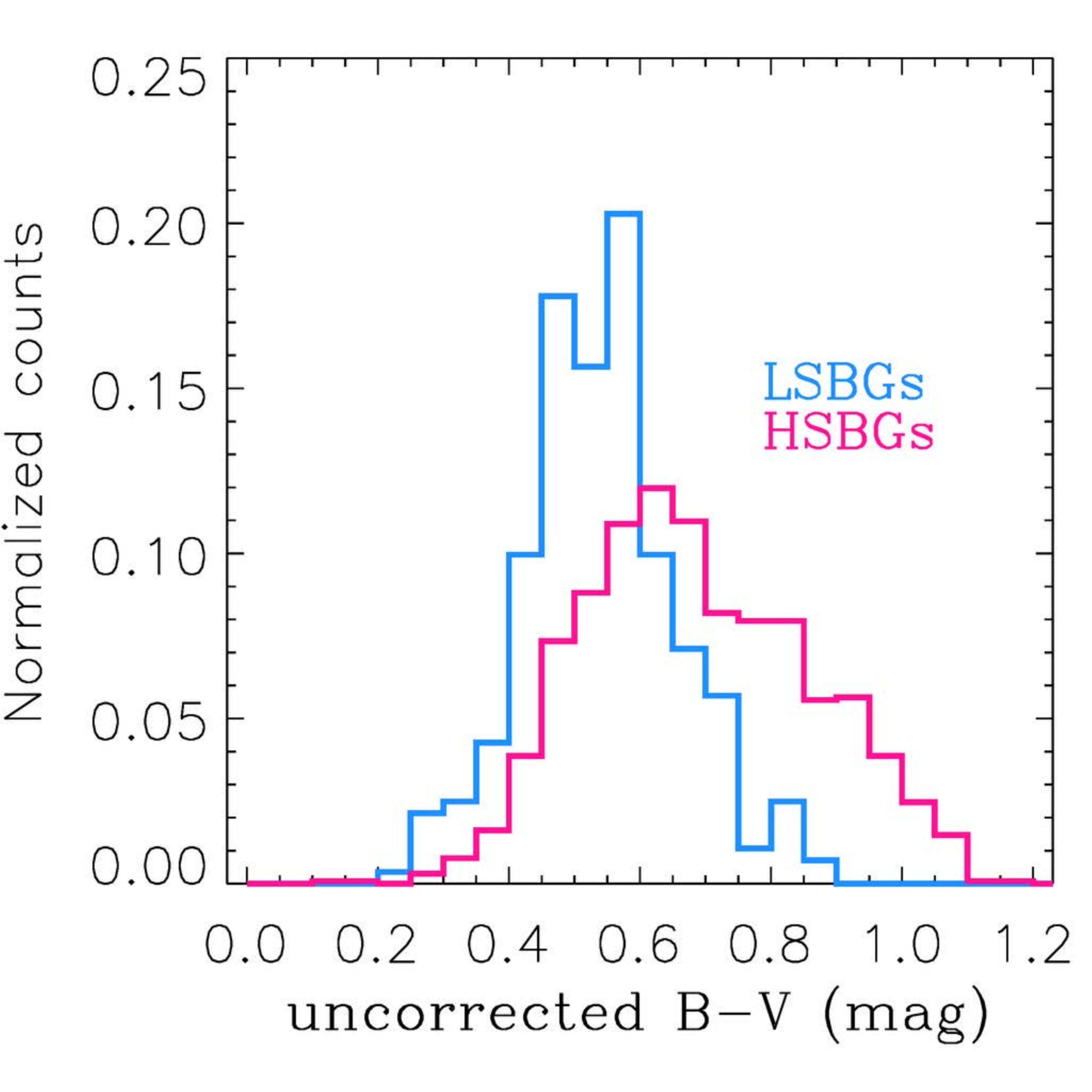}{0.25\textwidth}{(a)}
          \fig{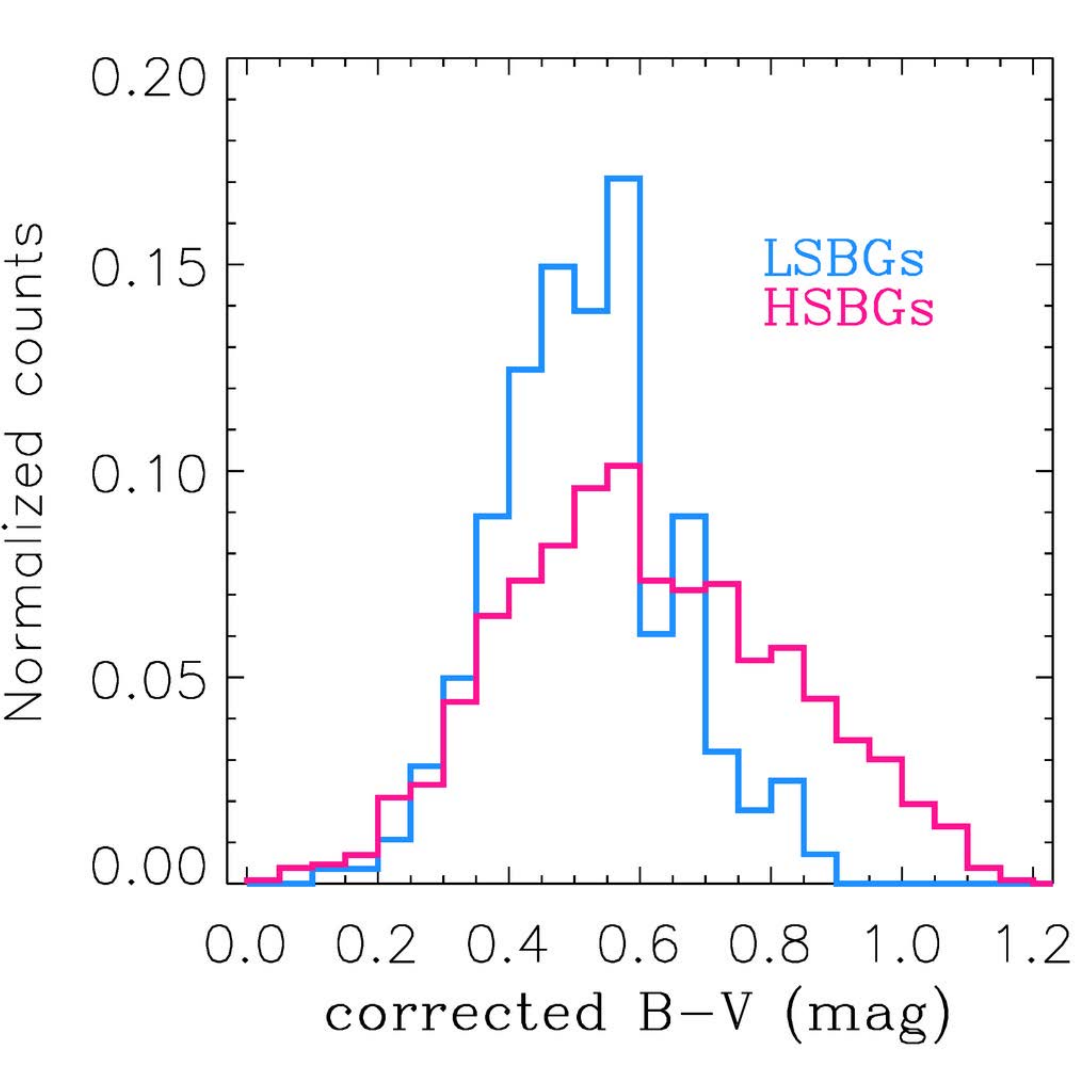}{0.25\textwidth}{(b)}
          \fig{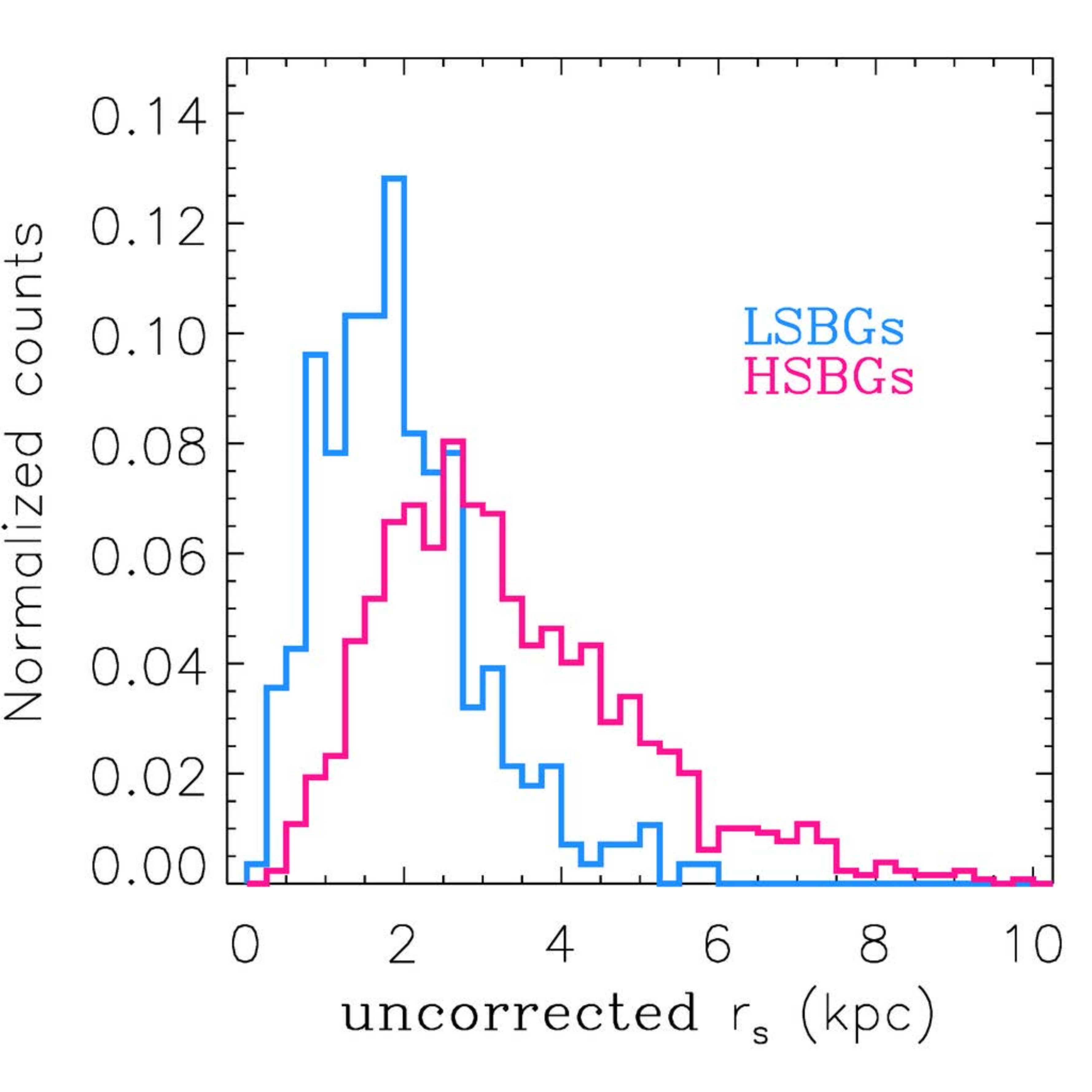}{0.25\textwidth}{(c)}
          \fig{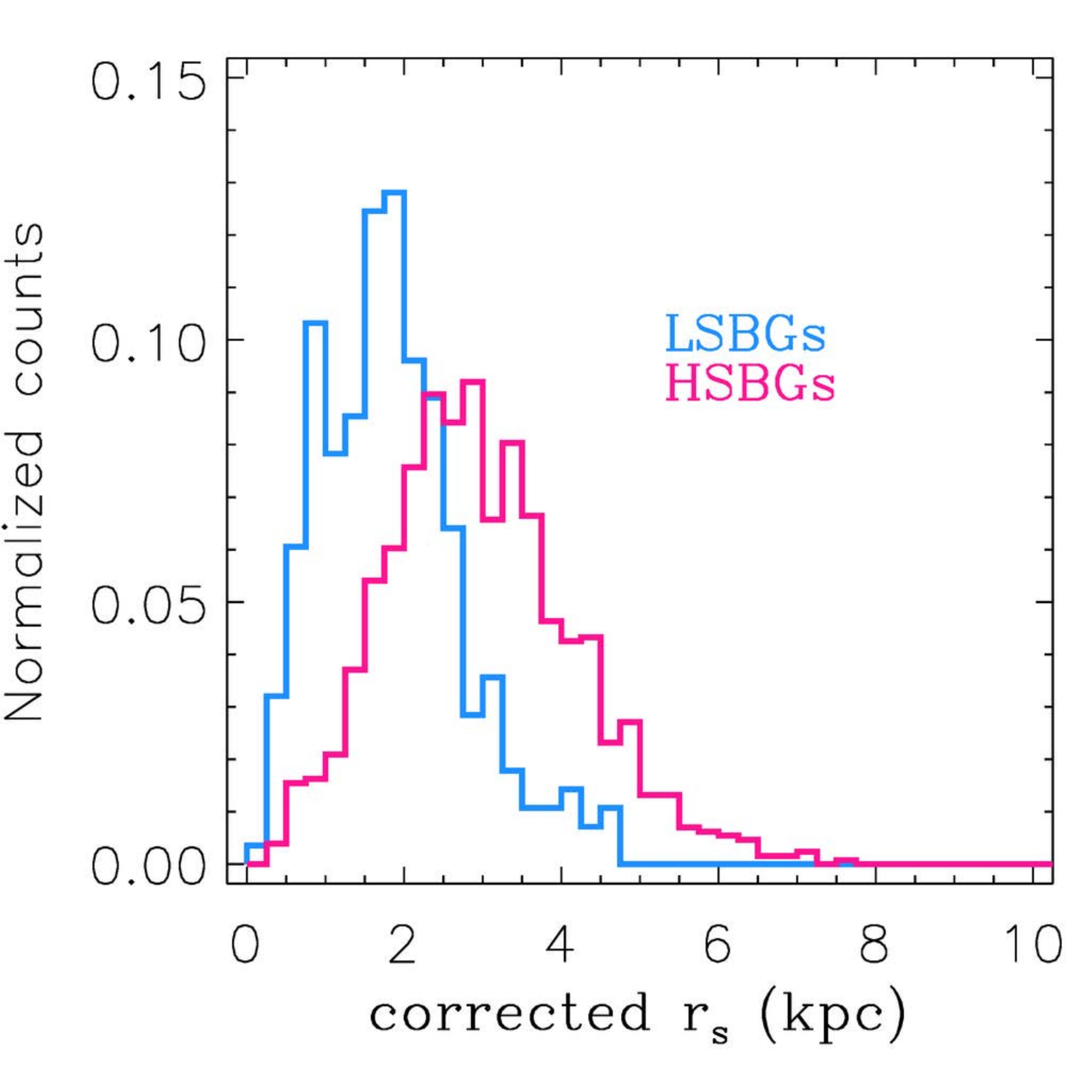}{0.25\textwidth}{(d)}
          }
\gridline{\fig{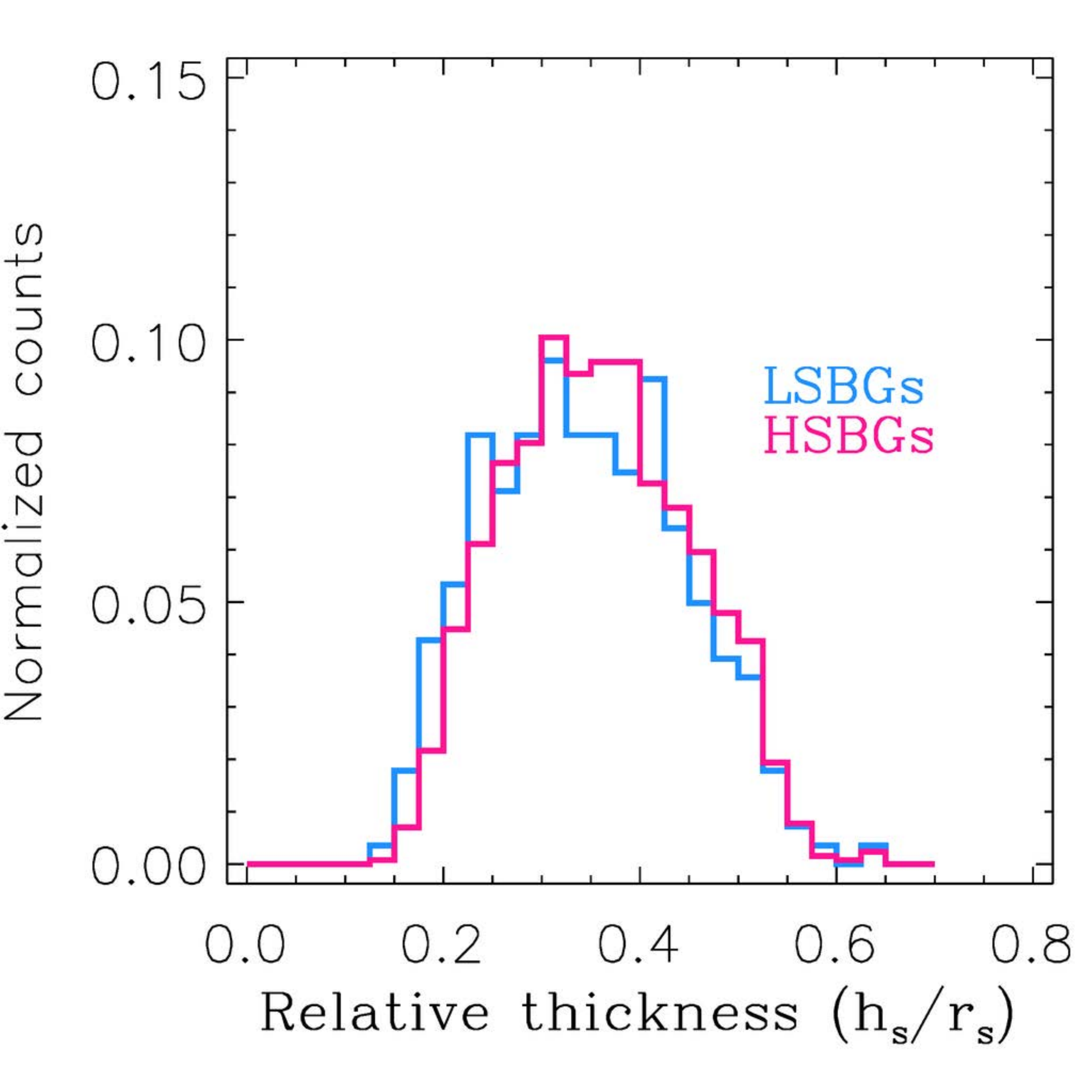}{0.25\textwidth}{(e)}
          \fig{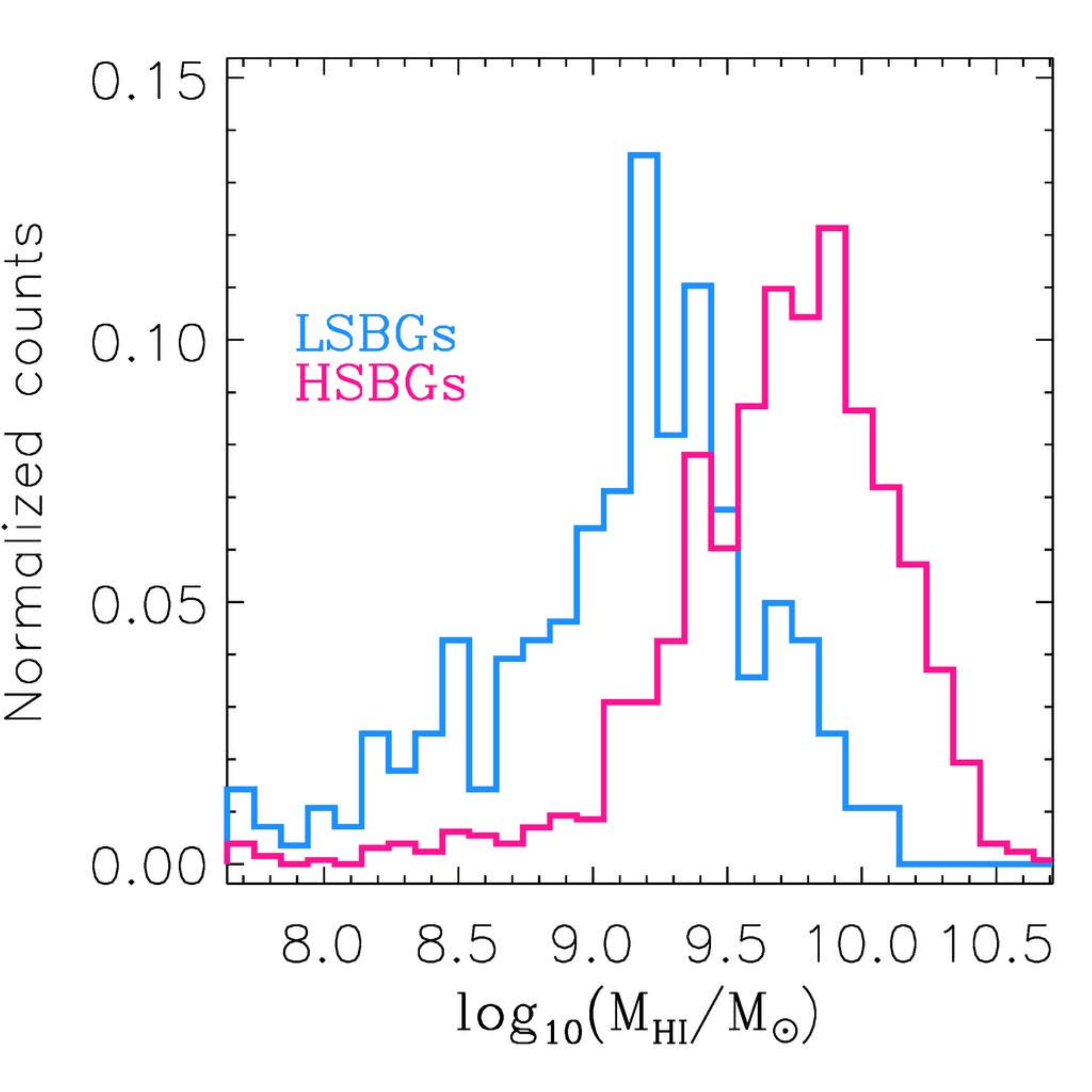}{0.25\textwidth}{(f)}
          \fig{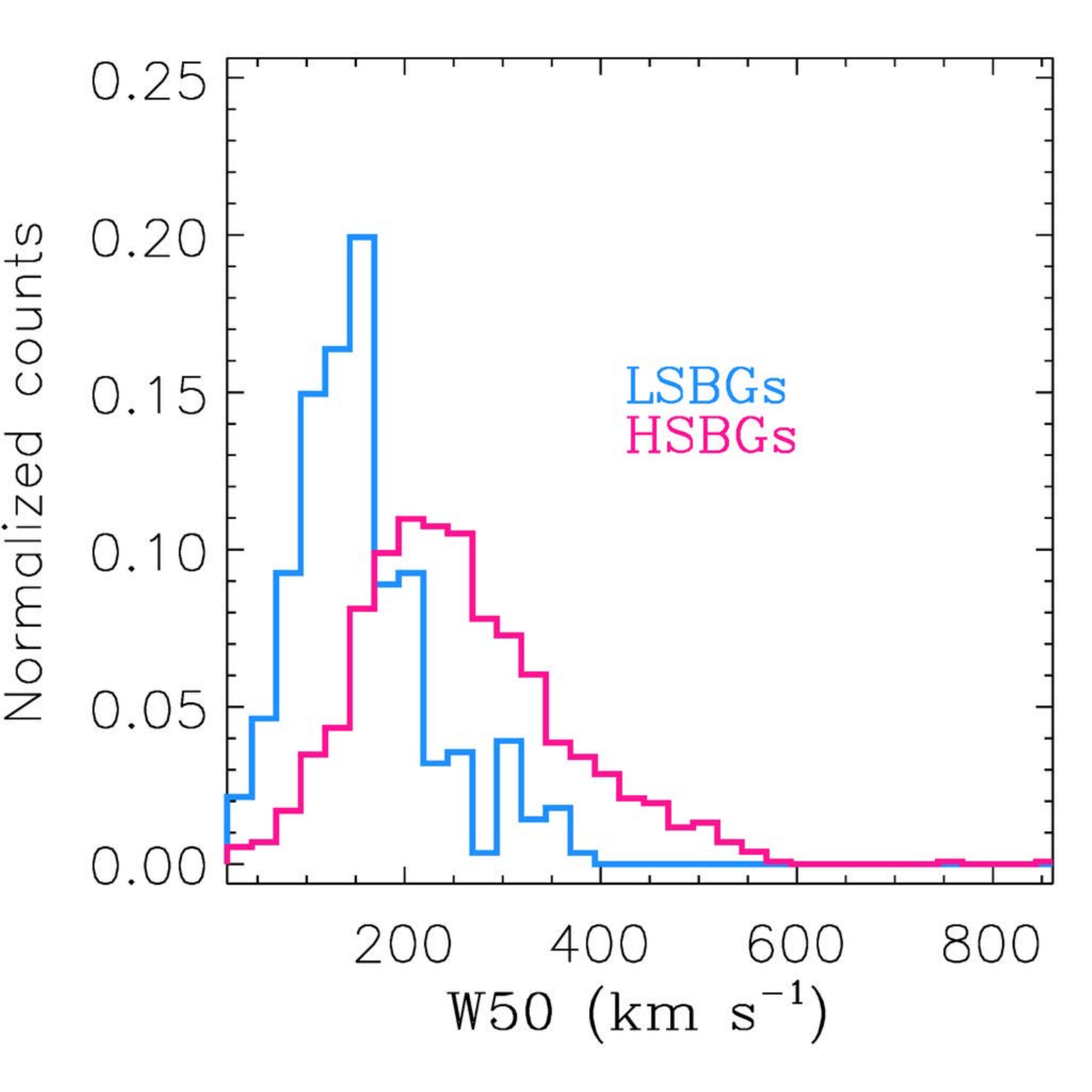}{0.25\textwidth}{(g)}
          }
\caption{These panels show the comparisons of the $B-V$ color, scale length $r_{\rm s}$, scale height $h_{\rm s}$ (before and after correcting internal extinction), H{\sc{i}}-mass and $W_{50}$ between our LSBG candidates (blue lines) and the remaining HSBGs (red lines) in our 1575 edge-on sample. The color of LSBG candidates are bluer than the HSBGs, and the $M_{\rm HI}$ and $W_{50}$ are lower than HSBGs. In the case of relative thickness, the $p$-value of K-S test for HSBGs and LSBGs is $13.0\%$, which indicates that LSBGs and HSBGs may be similar in the relative thickness.
\label{fig:his-L-HSBG}}
\end{figure*}

The comparison between the edge-on LSBG candidates and the remaining HSBGs in our edge-on sample is shown in Figure \ref{fig:his-L-HSBG}. Panels (a-b) show that the $B-V$ color of edge-on LSBG candidates are bluer than that of edge-on HSBGs, no matter whether correct the internal extinction or not. This conclusion is consistent with other literature of LSBGs \citep[e.g.,][]{Du2015}. According to Figure \ref{fig:his-L-HSBG}(f) and (g), our edge-on gas-rich LSBG candidates have lower H{\sc{i}} masses and smaller velocity widths compared to edge-on gas-rich HSBGs. All these panels show the significant differences between HSBGs and LSBGs, except Figure \ref{fig:his-L-HSBG}(e). The K-S test result for the relative thickness of HSBGs and LSBGs is $13.0\%$, indicates that LSBG and HSBG may be similar in the relative thickness.\\

\subsection{Absence of fainter LSBGs}
\label{subsec:absense}

We note that there are also few LSBGs at the very low-surface-brightness end, as exists in the faint tail of \citetalias{Du2015}'s sample (see Figure \ref{fig:properties}(d)). After checking images of these very faint galaxies, most of them are irregular galaxies, which is likely to have been rejected by the selection described in Section \ref{subsec:p-sample}.\\

\section{Summary}
\label{sec:summary}

In this work, we focus on selecting a sample of edge-on LSBGs from the $\alpha.40$ catalog. Firstly, do a preliminary selection of 1575 edge-on galaxies. Secondly, given that the estimations of central surface brightness, internal extinction and scale length would alter among different inclinations, we have corrected the inclination effects for these 1575 objects. Compared with face-on galaxies sample, the differences of the corrected absolute magnitude and corrected central surface brightness between our edge-on galaxies sample and face-on sample tend to be much smaller, which indicates the selection method adopted in this work is feasible.\\

Finally, we present a catalog of a sample of 281 edge-on LSBG candidates. Compared with the face-on LSBGs selected by \citepalias{Du2015}, our edge-on LSBG candidates prefer to have bluer color and fewer H{\sc{i}} gas than the face-on LSBGs. And there are more dwarf galaxies in our sample. Furthermore, a brief comparison between edge-on LSBGs and HSBGs shows that, in general, LSBGs present distinctive properties from HSBGs, including bluer color, smaller scale. But also represent a probably similarity of relative thickness between LSBGs and HSBGs. Except those, our result also suggests that inclination corrections are very important to obtain a complete sample of LSBGs.\\

\textbf{Acknowledgment} This project is supported by the National Key R$\&$D Program of China (No. 2017YFA0402704), the National Natural Science Foundation of China (NSFC) (Grant No. 11733006, No. 11403037) and the Key Laboratory of Optical Astronomy, National Astronomical Observatories, Chinese Academy of Sciences. W.D. is supported by the NSFC grant No. U1931109, and the Young Researcher Grant funded by National Astronomical Observatories, Chinese Academy of Sciences (NAOC). H.Y. acknowledges the National Natural Science Foundation of China (Grant No. 11803047). The authors sincerely thank Dr. Yan-Xia Zhang for her kindly help, thank the Sloan Digital Sky Survey project for providing their fpC-images for processing again, thank the ALFALFA project for providing the catalog of matched $\alpha.40$ and SDSS DR7 data.\\

\bibliography{edge-on}

\end{CJK*}
\end{document}